\documentclass{WileyMSP-template}
\usepackage{amssymb,multirow,slashbox}
\begin{document}

\pagestyle{fancy}

\title{Coherent Ising machines with error correction feedback}

\maketitle


\author{Satoshi Kako*}
\author{Timoth\'ee Leleu}
\author{Yoshitaka Inui}
\author{Farad Khoyratee}
\author{Sam Reifenstein}
\author{Yoshihisa Yamamoto}


\dedication{}

\begin{affiliations}
Satoshi Kako, Yoshitaka Inui, Farad Khoyrate, Sam Reifenstein, Yoshihisa Yamamoto\\
Physics \& Informatics Laboratories, NTT Research Inc.,\\
1950 University Ave., \#600, East Palo Alto, CA 94303, U.S.A.\\
Email Address:satoshi.kako@ntt-research.com\\

Timoth\'ee Leleu\\
Institute of Industrial Science, The University of Tokyo,\\
4-6-1 Komaba, Meguro-ku, Tokyo 153-8505, JAPAN\\
International Research Center for Neurointelligence, The University of Tokyo,\\
7-3-1 Hongo Bunkyo-ku, Tokyo 113-0033, JAPAN\\

\end{affiliations}



\begin{abstract}

A non-equilibrium open-dissipative neural network, such as a coherent Ising machine based on mutually coupled optical parametric oscillators, has been proposed and demonstrated as a novel computing machine for hard combinatorial optimization problems. However, there is a challenge in the previously proposed approach: The machine can be trapped by local minima which increases exponentially with a problem size. This leads to erroneous solutions rather than correct answers. In this paper, it is shown that it is possible to overcome this problem partially by introducing error detection and correction feedback mechanism. The proposed machine achieves efficient sampling of degenerate ground states and low-energy excited states via its inherent exploration property during a solution search process.

\end{abstract}


\section{Introduction}
Recently, a non-equilibrium open-dissipative artificial neural network consisting of optical\\ oscillators {\scriptsize  $^{\cite{Utsunomiya2011,Wang2013,Marandi2014,Inagaki2016,McMahon2016,Nixon2013,Takeda2017}}$} and Bose-Einstein condensates {\scriptsize $^{\cite{Berloff2017,Kalinin2018}}$} has been studied as a novel computing method for hard optimization problems. In those novel computing machines, the cost functions such as Ising Hamiltonian {\scriptsize $^{\cite{Utsunomiya2011,Wang2013,Marandi2014,Inagaki2016,McMahon2016}}$} and XY Hamiltonian {\scriptsize $^{\cite{Nixon2013,Takeda2017,Berloff2017,Kalinin2018}}$} are mapped to the loss landscape of the artificial neural network rather than the standard approach to map a target Hamiltonian to an energy landscape used in classical {\scriptsize $^{\cite{Kirkpatrick1983,Wang2015,Zarand2002}}$} and quantum annealing.{\scriptsize $^{\cite{Farhi2001,Kadowaki1998}}$} By providing an appropriate gain to such an open-dissipative artificial neural network with a slow enough speed, a lowest-loss ground state of a target Hamiltonian should be spontaneously selected as a single oscillation/condensation mode.{\scriptsize $^{\cite{Wang2013,Berloff2017,Leleu2017}}$}
\vspace{\baselineskip}

A unique advantage of using degenerate optical parametric oscillators (DOPOs) as neural nodes is its hybrid quantum and classical characters. At below threshold, the quantum noise correlation formed among DOPOs realizes a quantum parallel search to identify a ground state before sizable mean-fields build up in DOPOs, while the pitchfork bifurcation above threshold amplifies the amplitude of a selected ground state exponentially to form a deterministic (classical) computation result.{\scriptsize $^{\cite{Maruo2016,Yamamoto2017}}$} This particular system is referred to as coherent Ising machines (CIMs) in this paper. There is a challenge for CIM, which is a universal problem for any combinatorial optimizers including classical {\scriptsize $^{\cite{Kirkpatrick1983,Wang2015,Zarand2002}}$} and quantum annealing.{\scriptsize $^{\cite{Farhi2001,Kadowaki1998}}$} An exponentially many local minima easily trap a solver and make it to report a wrong answer when a problem size increases and yet a computational time is finite.{\scriptsize $^{\cite{Hamerly2019}}$} Recently, an error detection and correction feedback scheme has been proposed to overcome this problem in classical context.{\scriptsize $^{\cite{Leleu2019}}$} 
\vspace{\baselineskip}

In this paper, we extend the error detection and correction feedback technique discussed in ref. {\cite{Leleu2019}} to quantum domain. We show that by modulating a mutual coupling field, external pump rate and target intensity base on repeated energy measurements, the trapping in local minima can be suppressed, which leads to improved performance compared to an open-loop CIM without such error correction feedback.{\scriptsize $^{\cite{Utsunomiya2011,Wang2013,Marandi2014,Inagaki2016,McMahon2016}}$} Noise-free deamplification of the canonical coordinate $\hat{X}$ provided by degenerate parametric amplifying element with negative pump parameter $p$ ($<0$) plays an important role in destabilizing local minima. The DOPO quantum states stay close to minimum uncertainty states with amplitude squeezing rather than amplitude anti-squeezing during an entire search process. A saturation parameter is identified as a useful metric to quantify the ``quantumness'' of DOPOs. When a CIM consists of DOPOs with a large saturation parameter, the success probability of finding ground states is greatly improved for hard instances.

\begin{figure}
	\centering
	\includegraphics[scale=0.12]{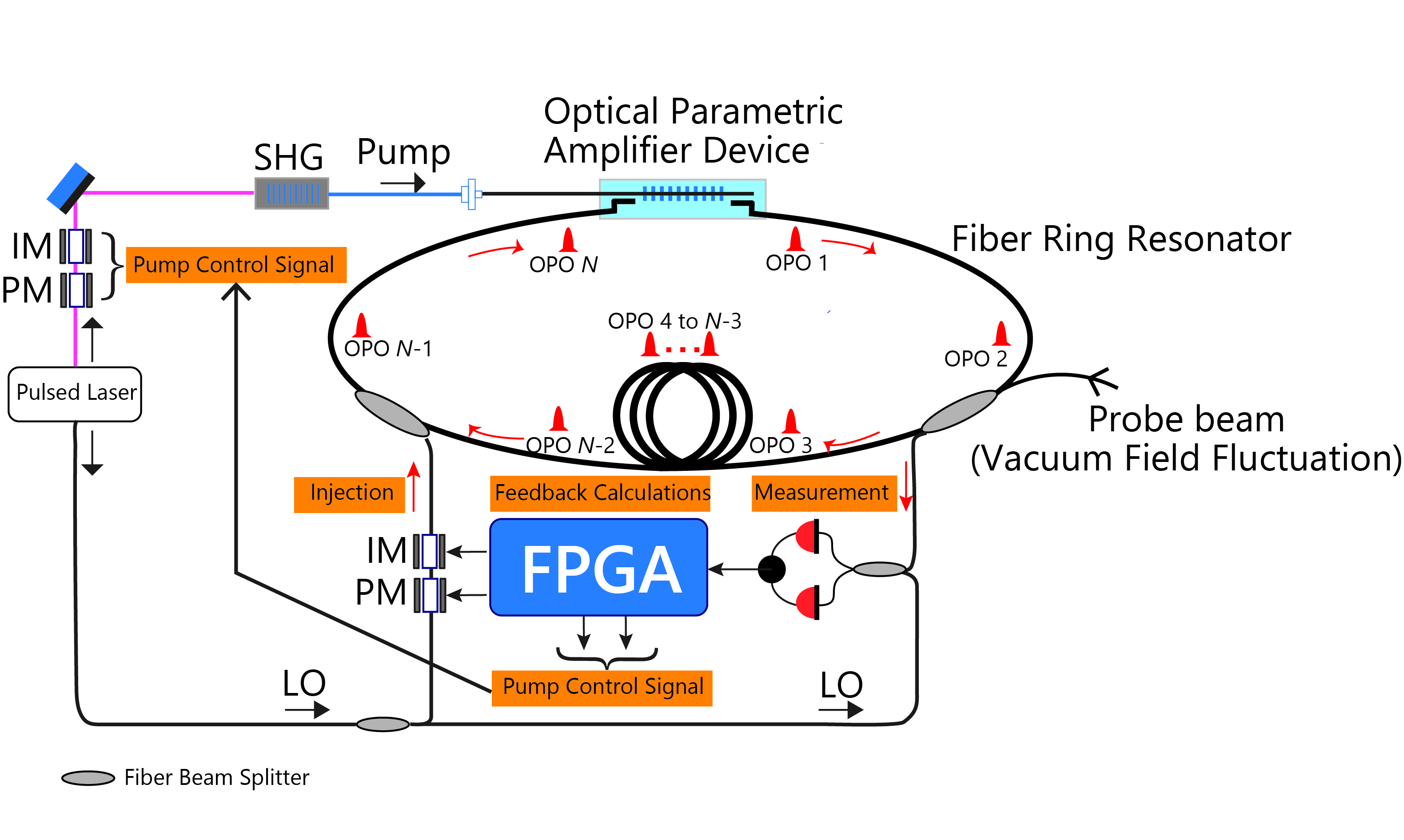}
	\caption{Schematic diagram of a proposed machine based on error detection and correction feedback.}
	\label{fig:setup}
\end{figure}

\section{Principle of the proposed machine}
The proposed machine is based on the measurement-feedback coupling CIM {\scriptsize $^{\cite{Inagaki2016,McMahon2016}}$} shown in figure \ref{fig:setup}. At each round trip in a ring resonator, every DOPO pulse in-phase amplitude (canonical coordinate) $\hat{X}_{i}$ ($i=1,\cdots, N$) is measured by an optical homodyne detector and a corresponding Ising spin is decided by the sign of a measured amplitude, that is, $S_{i}=\tilde{X}_{i}/|\tilde{X}_{i}|$, where $\tilde{X}_{i}$ is an inferred amplitude. Note that this is an indirect and weak measurement, in which a probe beam carries its own vacuum noise and only a small portion of the internal DOPO pulse field is extracted for measurement. If a current Ising energy $\mathcal{E}(t)=-\sum_{i<k} J_{ik} S_{i} S_{k}$, computed by $S_{i}$ ($i=1, \cdots, N$), is lower than the best Ising energy $\mathcal{E}_{opt}$ previously visited ($\mathcal{E}(t) < \mathcal{E}_{opt}$), we increase the DOPO pump rate ($p>0$) and simultaneously increase the mutual coupling field ($e>0$) among DOPOs in order to decrease an energy continuously by flipping ``wrong spins'' and preserving ``correct  spins''. This mode of operation is similar to that of an open-loop CIM, in which the pump rate $p$ is monotonically increased  {\scriptsize $^{\cite{Inagaki2016}}$} or the mutual coupling field $e$ is monotonically increased from below to above threshold.{\scriptsize $^{\cite{McMahon2016}}$} This step is called a mode ``A''. If a current Ising energy $\mathcal{E}(t)$ is roughly equal to the best Ising energy $\mathcal{E}_{opt}$, we assume the machine already visited a local minimum. We then decrease the pump rate $p$ close to zero ($p\simeq0$) and eliminate the central barrier of the effective potential, $V(X)=1/2(1-p)X^2 + (1/4)g^2 X^4$, to let all Ising spins to switch freely by the mutual coupling field $e$. In this way, we can avoid the notorious problem of getting trapped in a local minimum. This mode of operation is called a mode ``B''. On the other hand, if a current Ising energy is higher than the best Ising energy previously visited ($\mathcal{E}(t) > \mathcal{E}_{opt}$), we assume the machine already started to escape from a local minimum by climbing up a potential. We then decrease the DOPO pump rate $p$ to a negative value ($p<0$) to destabilize the current spin configuration more strongly and simultaneously maintain the high level of mutual coupling field among DOPOs in order to identify which spins should be flipped and which other spins should be maintained to move away from the previously visited local minimum. This mode of operation is called a mode ``C''. 
\vspace{\baselineskip}

\begin{figure}
	\centering
	\includegraphics[scale=0.12]{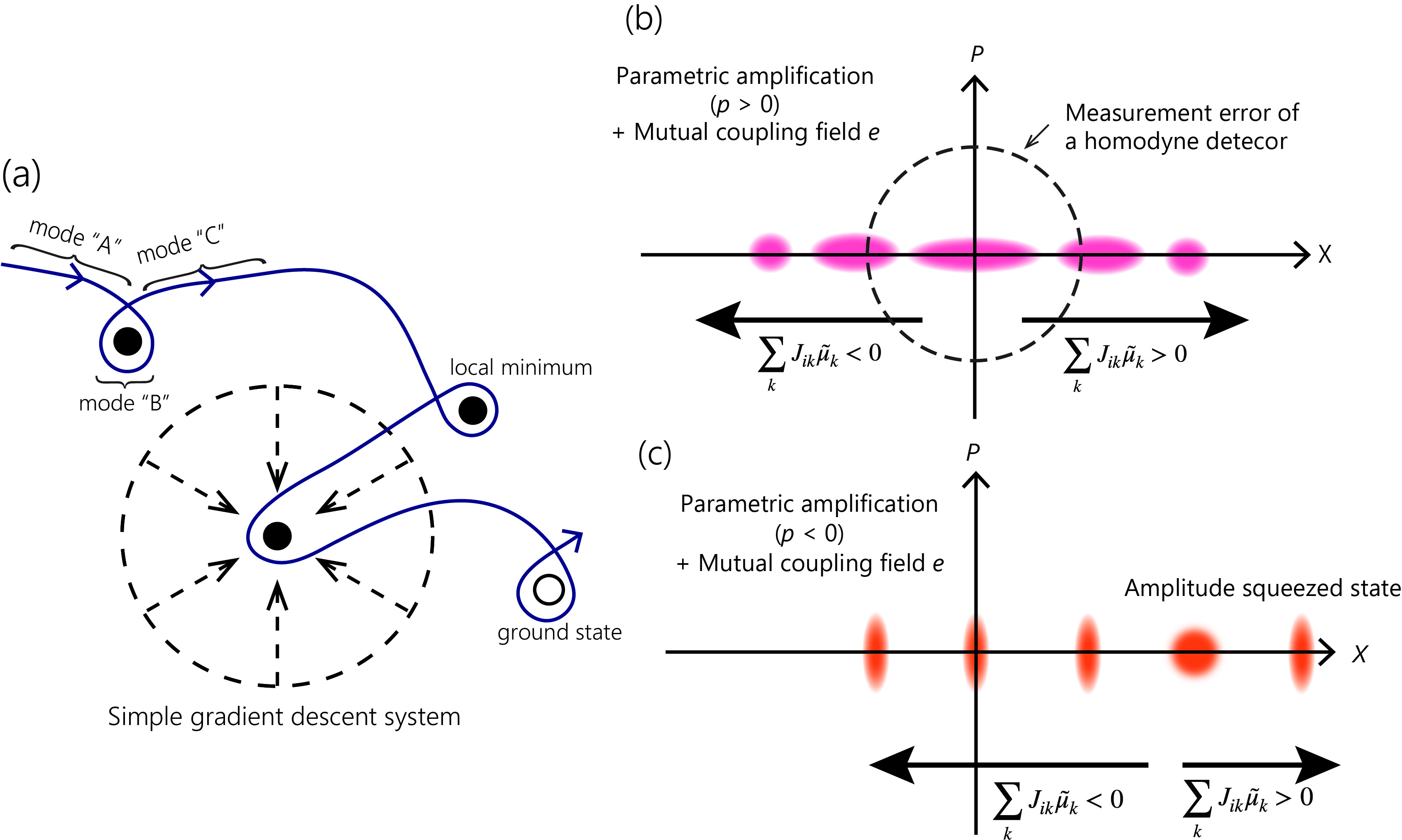}
	\caption{(a) A system trajectory of CIM with error correction feedback (solid line) and a simple gradient descent system (dashed line). (b) An open-loop CIM employs parametric amplification to search for a ground state. A measurement error is on the order of $w/\sqrt{R_B}$, where $w$ is a vacuum fluctuation and $R_B$ is the reflectivity of an out-coupling beam splitter. (c) A closed-loop CIM with error correction feedback employs parametric deamplification to search for a ground state.} 
	\label{fig:op-principle}
\end{figure}

As a consequence of such dynamical modulation of the pump rate $p$ and the mutual coupling field $e(t)$, the system trajectory avoids getting trapped in a local minimum and migrates from one local minimum to another in search of a true ground state, as shown in figure \ref{fig:op-principle}(a). The above three operational modes (A, B, C) repeat when the machine explores over many local minima. On the contrary, a simple gradient decent system relaxes to a particular local minimum determined by an initial condition as shown by the dashed lines in figure \ref{fig:op-principle}(a). 
\vspace{\baselineskip}

In an open-loop CIM with an increasing pump rate {\scriptsize $^{\cite{Inagaki2016}}$} or mutual coupling field,{\scriptsize $^{\cite{McMahon2016}}$} a parametric amplification gain and gain saturation (two photon absorption) make the state of each DOPO pulse evolve from an anti-squeezed vacuum state to a coherent state, for which the signal-to-noise ratio including detector noise is improved because of an increased mean-field, reduced quantum noise and constant measurement error, as shown in figure \ref{fig:op-principle}(b). In a closed-loop CIM with error detection and correction feedback, on the other hand, the parametric deamplification is switched on ($p<0$) during a mode ``C'', so that the quantum states of both spin flipping and spin preserving DOPO pulses become amplitude squeezed states, in which the mutual coupling field $e$ decides which spins are flipped and which other spins are preserved, as shown in figure \ref{fig:op-principle}(c). A negative pump rate ($p<0$) exponentially decreases the absolute amplitude $|\langle \hat {X} \rangle |$, which helps DOPOs to switch the polarity from positive to negative, and vice versa, quickly. Since the negative pump rate squeezes the amplitude noise $\langle \Delta \hat{X}^{2}\rangle$ when DOPOs cross a zero point, those switching DOPO pulses are more sensitive to a mutual coupling field $e$.

\section{Gaussian quantum theory}
The proposed machine is modeled by the quantum mechanical master equation with three Liouvillian coupling terms for dissipative coupling among DOPO pulses aided by a measurement-feedback circuit, two photon absorption loss (or back conversion from signal to pump fields) in a degenerate parametric amplifying device, and background linear loss, respectively.{\scriptsize $^{\cite{Yamamura2017, Shoji2017}}$} By expanding the field density operator with the Wigner distribution function, we can obtain the Fokker-Planck equation for a signal field variable after adiabatic elimination of a pump-field variable and appropriate truncation of higher-order derivative terms. Then, we derive the truncated-Wigner stochastic differential equation (W-SDE) using the Ito rule.{\scriptsize $^{\cite{Maruo2016}}$} Validation of this approach (W-SDE) was confirmed by comparing the entanglement and quantum discord computed by this model with those computed by more accurate positive-P stochastic differential equation (P-SDE) for an optical delay-line coupling CIM.{\scriptsize $^{\cite{Takata2015,Inui2019,Inui}}$}
\vspace{\baselineskip}

A solitary DOPO is described by the parametric interaction Hamiltonian $\hat{\mathcal H} = i \hbar \frac{p}{2} (\hat{a}^{+2} - \hat{a}^2)$, the projector for a single photon loss $\hat{L}_{1} = \hat{a}$ and that for a two photon loss $\hat{L}_{2} = \sqrt{\frac{g^2}{2}} \hat{a}^2$. Here $g^2$ is a saturation parameter and equal to the inverse photon number at twice threshold pump rate. In the case of a small saturation parameter, $g^2 \ll 1$, we can separate the i-th DOPO pulse amplitude to the mean field  and small fluctuation, $\hat{X}_i = \langle \hat{X}_i\rangle + \Delta \hat{X}_i$. The equation of motion for the mean field $\mu_i = \langle \hat{X}_i\rangle /2$ is (see Appendix A),
\begin{equation}{\label{eq:mu-equation}}
\frac{d}{dt} \mu_i =\left[ - \left( 1 + j \right) + p - g^2 \mu^2_i \right] \mu_i + \frac{e_i (t)}{\sqrt{{\frac{1}{N}} \sum_{i,j} \left| J_{ij} \right|}} \sum_{k}{J_{i k} \left( j \mu_k + \sqrt{\frac{j}{4}} w_k \right)} + \sqrt{j} \langle :\Delta \hat{X^2_i}: \rangle w_i.
\end{equation}
Here a time $t$ is normalized so that the background linear loss (amplitude decay rate) is one in this time unit, as indicated in the first term of the R.H.S. of equation (\ref{eq:mu-equation}). Another loss parameter $j$ represents the normalized out-coupling rate for optical homodyne measurement, where $j=R_B/\Delta t_c$, $R_B$ is the reflectivity of the out-coupling beam splitter and $\Delta t_c$ is the round trip time of a ring cavity. We assume $R_B = {j \Delta t}_c \ll 1$. $p$ is a linear gain coefficient provided by the parametric device. The term $g^2 \mu^2_i$ expresses two photon absorption rate (back conversion rate from signal to pump fields). A solitary DOPO, without mutual coupling ($J_{ik} = 0$) and without out-coupling loss for measurement ($j=0$), has an oscillation threshold $p_{th} = 1$ and an average photon number at above threshold $\langle \hat{n}_i \rangle \simeq {\mu_i}^2  = (p-1)/g^2$. $w_k$ is a zero-mean and variance-one real number Gaussian random variable, which accounts for a finite measurement uncertainty in optical homodyne detection and is mainly determined by a vacuum field fluctuation incident upon the open port of the out-coupling beam splitter (XBS in figure \ref{fig:setup}). An inferred mean-field amplitude, $\tilde{\mu}_k = \mu_k + \sqrt{\frac{1}{4j}} w_k$, is deviated from the true mean amplitude $\mu_k$ by the finite measurement uncertainty $\sqrt{\frac{1}{4j}} w_k$. $J_{ik}$ is the Ising coupling coefficient and $e(t)$ is a dynamically modulated feedback mean-field. $\langle :\Delta \hat{X}^2_i: \rangle = \langle \Delta \hat{X}^2_i \rangle - 1/2$ is a normally ordered variance which represents excess amplitude noise above the standard quantum limit ${\langle \Delta \hat{X}^2 \rangle}_{SQL} = 1/2$. The second and third terms of the R.H.S. of equation (\ref{eq:mu-equation}) represent the (noisy) measurement feedback coupling term and the measurement-induced shift of the mean-field, respectively. Note that the measurement-induced shift of DOPO wavepacket disappears when an internal field is in a coherent state with $\langle :\Delta \hat{X}^2_{i}:\rangle = 0$, because there is no entanglement between the internal and external fields in this special case.
\vspace{\baselineskip}

The equation of motion for the variance $\sigma_i = \langle \Delta \hat{X}^2_i \rangle$ is obtained as (see Appendix A)
\begin{equation}{\label{eq:val-equation}}
\frac{d}{dt} \sigma_i = \frac{d}{dt} \langle \Delta {\hat{X}_i}^2 \rangle = 2 \left[ - \left( 1 + j \right) + p - 3 g^2 \mu^2_i \right] \sigma_i - 2j {\left( \sigma_i -1/2 \right)}^2 + \left[ \left( 1 + j \right) + 2 g^2 \mu^2_i \right].
\end{equation}
The first term of the R.H.S. of equation (\ref{eq:val-equation}) manifests that the variance $\sigma_i$ is attenuated by linear loss, amplified by parametric gain, and attenuated by two-photon absorption loss. The second term of the R.H.S. of equation (\ref{eq:val-equation}) represents the measurement-induced reduction of the DOPO quantum state. Note that there is no state reduction if the internal DOPO pulse is in a coherent state ($\sigma_i = 1/2$), for which there is no quantum correlation between the internal DOPO pulse and the out-coupled pulse for measurement so that there is no back action imposed on the internal DOPO quantum state by the measurement. The third term of the R.H.S. of equation (\ref{eq:val-equation}) shows that the variance increases by the incident vacuum field fluctuation via linear loss and by the pump noise via gain saturation, respectively. The equation of motion for the variance $\eta_i=\langle \Delta {\hat{P}_i}^2 \rangle$ is similarly obtained as, {\scriptsize $^{\cite{Inui}}$}
\begin{equation}{\label{eq:Pval-equation}}
\frac{d}{dt} \eta_i = \frac{d}{dt} {\langle \Delta {\hat{P}_i}^2 \rangle} = 2 \left[ - \left( 1 + j \right) + p - g^2 \mu^2_i \right] \eta_i + \left[ \left( 1 + j \right) + 2 g^2 \mu^2_i \right].
\end{equation}
Note that this equation is decoupled from the equations for $\mu$ and $\sigma$ so that we do not need to solve it to search for the solutions of combinatorial optimization problems. However, it is worth computing in order to understand the quantum property of the proposed CIM as discussed in the following sections.   
\vspace{\baselineskip}

When there is no external pumping ($p=0$), equations (\ref{eq:mu-equation}), (\ref{eq:val-equation}), and (\ref{eq:Pval-equation}) show that there exists no mean-field $\mu_i=0$ but there are finite variance $\sigma_i= \eta_i = 1/2$. This is the vacuum field noise injected constantly from zero temperature reservoirs. When the pump rate is far above threshold ($p \gg (1+j)$), the mean-field is $\mu_i = \pm \sqrt{p/g^2}$ and the variance is approaching to $\sigma_i = \mu_{i} = 1/2$. This is the quantum noise associated with a coherent state produced in a highly excited DOPO.
\vspace{\baselineskip}

The dynamically modulated feedback mean-field $e_i (t)$ obeys the following equation, {\scriptsize $^{\cite{Leleu2019}}$}
\begin{equation}{\label{eq:error-equation}}
\frac{d}{dt} e_i (t) = - \beta \left[ g^2 {\tilde{\mu}_i}^2 - a \right] e_i (t).
\end{equation}
Here $\beta$ is a positive constant and $a$ is the target intensity (squared amplitude). Note that the feedback mean-field is exponentially increased if the normalized inferred intensity $g^2 {\tilde{\mu}_i}^2$ is smaller than the target intensity $a$, while it is exponentially decreased if the opposite is true, that is, $g^2 {\tilde{\mu}_i}^2 > a$. $\beta$ is considered as a rate to reach a steady state condition $g^2 {\tilde{\mu}_i}^2= a$. Finally, the pump rate $p$ and target intensity $a$ are determined by the difference between the current Ising energy $\mathcal{E}(t)$ and the best Ising energy $\mathcal{E}_{opt}$ visited before: 
\begin{eqnarray}
	{\label{eq:p}}
	p(t) &=& \pi - \rho_{p} \tanh \left( \frac{\mathcal{E}(t) - \mathcal{E}_{opt}}{\Delta} \right) \\
	{\label{eq:a}}
	a(t) &=& \alpha + \rho_{a} \tanh \left( \frac{\mathcal{E}(t) - \mathcal{E}_{opt}}{\Delta} \right)
\end{eqnarray}
Here $\pi$, $\alpha$, $\rho_p$, $\rho_a$, and $\Delta$ are positive constants, and $\tanh (x)$ is a hyperbolic tangent function. If $\pi < \rho$, the pump rate $p$ becomes negative when $\mathcal{E}(t) - \mathcal{E}_{opt} \gg \Delta$ (mode ``C''), while $p$ is positive when the opposite is true (mode ``A''), as described already in the previous section.

\begin{figure}[h]
	\centering
	\includegraphics[scale=0.38]{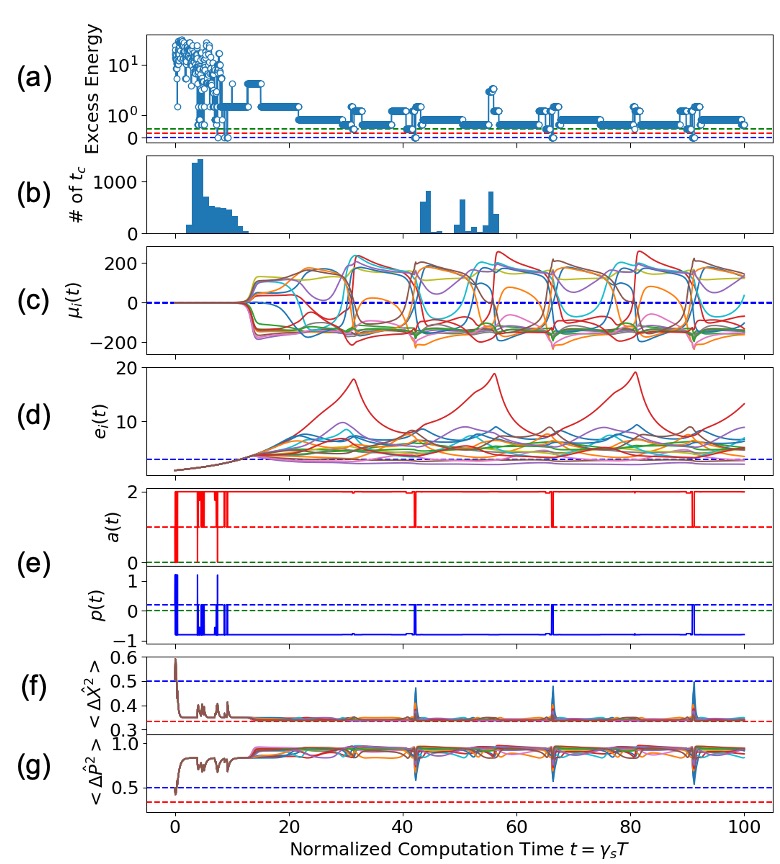}
	\caption{Dynamical behavior of CIM with error correction feedback. (a) Ising energy, (b) Histogram of evolution time $t_s$ finding a ground state for the first time, (c) Mean-field amplitude $\mu(t)$, (d) Mutual coupling field $e(t)$, (e) Pump rate $p(t)$ and target intensity $a(t)$, (f) Variance in canonical coordinate $\langle \Delta \hat{X}^2 \rangle$ and $(g)$ Variance in canonical momentum $\langle \Delta \hat{P}^2 \rangle$. The feedback parameters are $\alpha = 1.0$, $\pi = 0.1$, $\rho_a = \rho_p = 1.0$, $\Delta = 1/5$ and the gain saturation parameter $g^2 = 10^{-4}$.}
	\label{fig:time response}
\end{figure}

\section{Numerical Simulation}
\subsection{Dynamical behavior of the machine}
We solve MAX-CUT problems with randomly chosen 21-level discrete weights $J_{ik}=(-1, -0.9,\cdots, 0.9, 1)$, for which an exact solution with a lowest energy is obtained by brute force search. Figure \ref{fig:time response} shows \textbf{(a)} the dynamical behavior of an inferred Ising energy $\mathcal{E}(t)$, \textbf{(b)} histogram of evolution time $t_s$ when one of exact solutions is obtained for the first time for 10,000 independent runs, \textbf{(c)} mean-amplitude $\mu(t)$, \textbf{(d)} feedback mean-field $e(t)$, \textbf{(e)} pump rate $p(t)$ (in blue) and target intensity $a(t)$ (in red), \textbf{(f)} canonical coordinate variance $\langle \Delta \hat{X}^2 \rangle$ and \textbf{(g)} canonical momentum variance $\langle \Delta \hat{P}^2 \rangle$. An evolution time is normalized by a linear loss rate $\gamma_s$, i.e. $t = \gamma_s T$, where $T$ is a wall clock time. The results shown in figure \ref{fig:time response} is a single run trajectory of the machine for a particular problem instance, except for the histogram of evolution time $t_s$ shown in the second panel (figure \ref{fig:time response}(b)) which is the result of $10,000$ trials. A problem size (number of spins) of the instance is $N = 16$. The feedback parameters are set to $\alpha = 1.0$, $\pi = 0.2$, $\rho_a = \rho_p = 1.0$, $\Delta = 1/5$ and $\beta = 0.05$. The saturation parameter is $g^2 = 10^{-4}$. We also assume that a signal field lifetime $\tau_{s}=1/\gamma_s$ is 40 times of a round trip time, $\gamma_s \Delta t_c = 0.025$, which corresponds to the reflectivity of the out-coupling beam splitter $R_B = 0.025$. In Appendix B, we discuss the difference in mean amplitudes, $\mu(t)$, in the closed-loop CIM and open-loop CIM.
\vspace{\baselineskip}

\begin{figure}[h]
	\centering
	\includegraphics[scale=0.1]{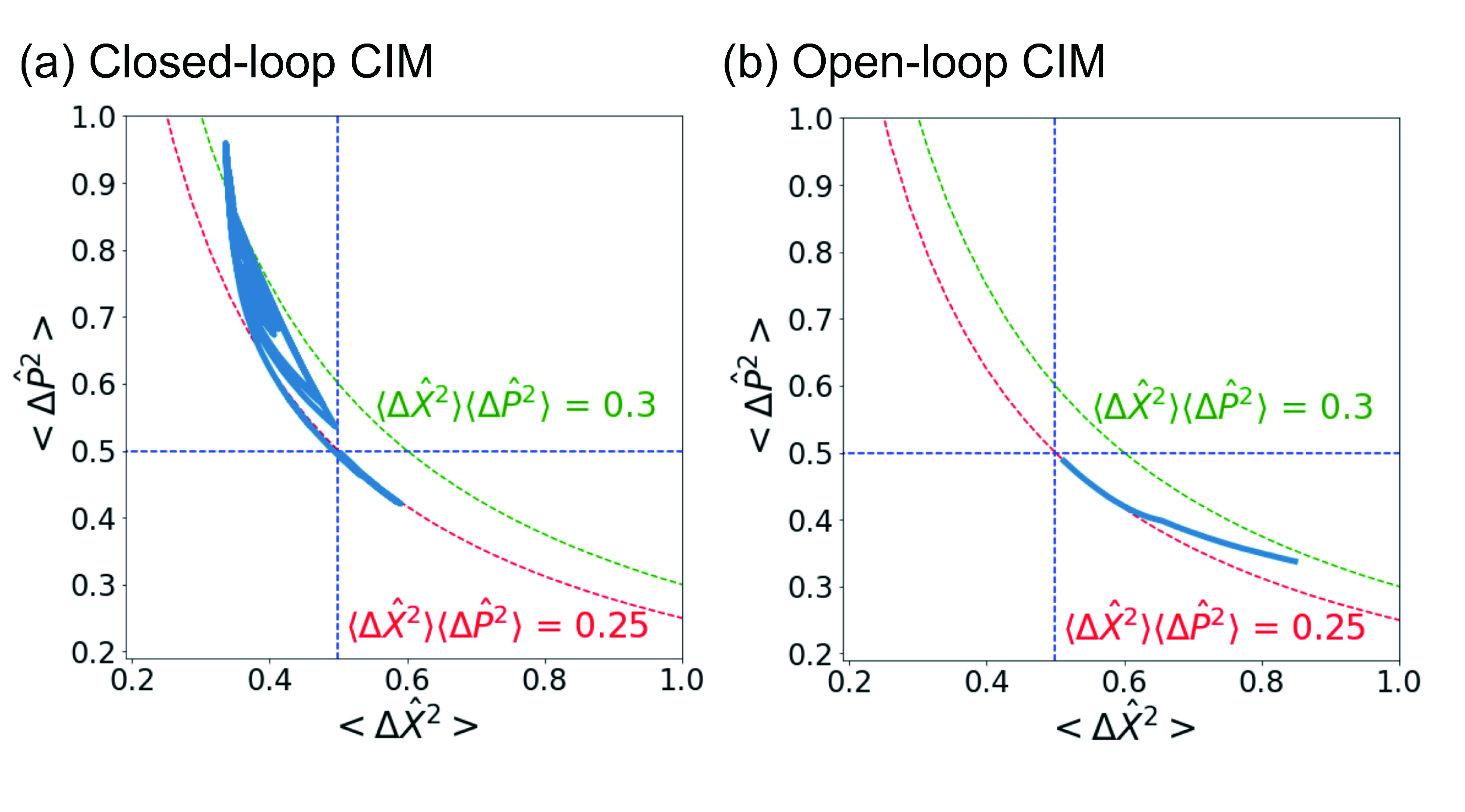}
	\caption{Variances $\langle \Delta \hat{X}^2 \rangle$ and $\langle \Delta \hat{P}^2 \rangle$ for (a) closed-loop CIM and (b) open-loop CIM. Numerical parameters are same as in figure \ref{fig:time response}}
	\label{fig:uncertainty}
\end{figure}

As shown in figure \ref{fig:time response}(a), the inferred Ising energy $\mathcal{E}(t)$ fluctuates rapidly between $t = 0$ and $t=12$, but settles down near the optimum energy $\mathcal{E}_{opt}$ by $t = 12$, when the average photon number of each DOPO exceeds one ($\mu_i^2 > 1$). For more than 50\% of 10,000 trials, the machine finds an exact solution before reaching this point (first bunch of the histogram between $t = 0$ and $t = 12$ in figure \ref{fig:time response}(b)). This is the situation that the machine finds exact solutions only after visiting a small number of local minima. Note that the machine repeats the modes ``A'', ``B'', ``C'' many times already between $t = 0$ and $t=12$, as a closed-loop CIM. This can be easily confirmed by the frequent switching of a pump rate $p(t)$ across $P=0$ as shown in figure \ref{fig:time response}(e). For the rest of 10,000 trials ($< 50\%$), the second bunch of success histogram is observed between $t=40$ and $t =60$, for which the machine needs to visit and escape from many local minima until it finally finds an exact solution. As shown in figure \ref{fig:time response}(d), the feedback mean-field $e(t)$ initially increases exponentially and then saturates when the actual DOPO intensity $g^2 \tilde{\mu}^2_{i}$ becomes equal to the target intensity $a(t)$. The pump rate $p(t)$ fluctuates rapidly between $p=1.2$ and $p=-0.8$ initially (between $t=0$ and $t=12$), but after $t=12$ the pump rate is set to a negative value ($p = - 0.8$) for most of time as shown in figure \ref{fig:time response}(e). This result indicates the instantaneous energy $\mathcal{E}(t)$ is larger than the best energy $\mathcal{E}_{opt}$ previously visited most of time, so that the machine is mostly in the mode ``C''. Several spins are flipped simultaneously with some intervals. At specific times $t = 42, 66, 91$ (almost periodically), $p(t)$ approaches $p = \pi = 0.2$ and $a(t)$ approaches $a = \alpha = 1$, which means the instantaneous energy $\mathcal{E}(t)$ becomes nearly equal to $\mathcal{E}_{opt}$ (see equations(\ref{eq:p}) and (\ref{eq:a})). The machine is close to a local minimum at those times. The flipping of many spins is observed at those times as shown in figure \ref{fig:time response}(c), which indicates that the machine already tries to escape from this local minimum. 
\vspace{\baselineskip}

Finally figure \ref{fig:time response} (f) and (g) show that the canonical coordinate $\hat{X}$ is squeezed ($\langle \Delta \hat{X}^2 \rangle < 1/2$) while the canonical momentum $\hat{P}$ is anti-squeezed ($\langle \Delta \hat{P}^2 \rangle > 1/2$) during the mode ``C'', as discussed in the previous section. This is in sharp contrast to a standard open-loop CIM, in which the amplitude anti-squeezing, $\langle \Delta \hat{X}^2 \rangle > 1/2$, and the phase squeezing, $\langle \Delta \hat{P}^2 \rangle < 1/2$, are maintained all the time.{\scriptsize $^{\cite{Yamamoto2017}}$} It is noted that the quantum states of all DOPO pulses satisfy the minimum uncertainty product, $\langle \Delta \hat{X}^2 \rangle \langle \Delta \hat{P}^2 \rangle = 1/4$, with a very small excess factor of less than 30\% for both machines, as shown in figure \ref{fig:uncertainty}. This result suggests that DOPO quantum states remain nearly pure states during entire computation time in spite of an open-dissipative nature of the machine. These results are distinct from optical delay line coupling CIM, in which the uncertainty product $\langle \Delta \hat{X}^2 \rangle \langle \Delta \hat{P}^2 \rangle$ is much greater than the minimum value.{\scriptsize $^{\cite{Inui}}$} Repeated weak quantum measurements in MFB-CIM contribute to the collapse the DOPO wavepacket to a minimum uncertainty state and maintain the purity of states.
\vspace{\baselineskip}

It is seen from figure \ref{fig:uncertainty}(a) that a closed-loop CIM employs a parametric amplification ($p > 0$) only in an initial transient but utilizes a parametric deamplification ($p < 0$) in a later time. The trajectory of a closed-loop CIM never converges to a specific state but continues to explore. On the other hand, an open-loop CIM always employs a parametric amplification. The trajectory of an open-loop CIM starts from a vacuum state, reaches a maximally anti-squeezed vacuum state at threshold and approaches to a coherent state toward the end of computation.

\begin{figure}[h]
	\centering
	\includegraphics[scale=0.18]{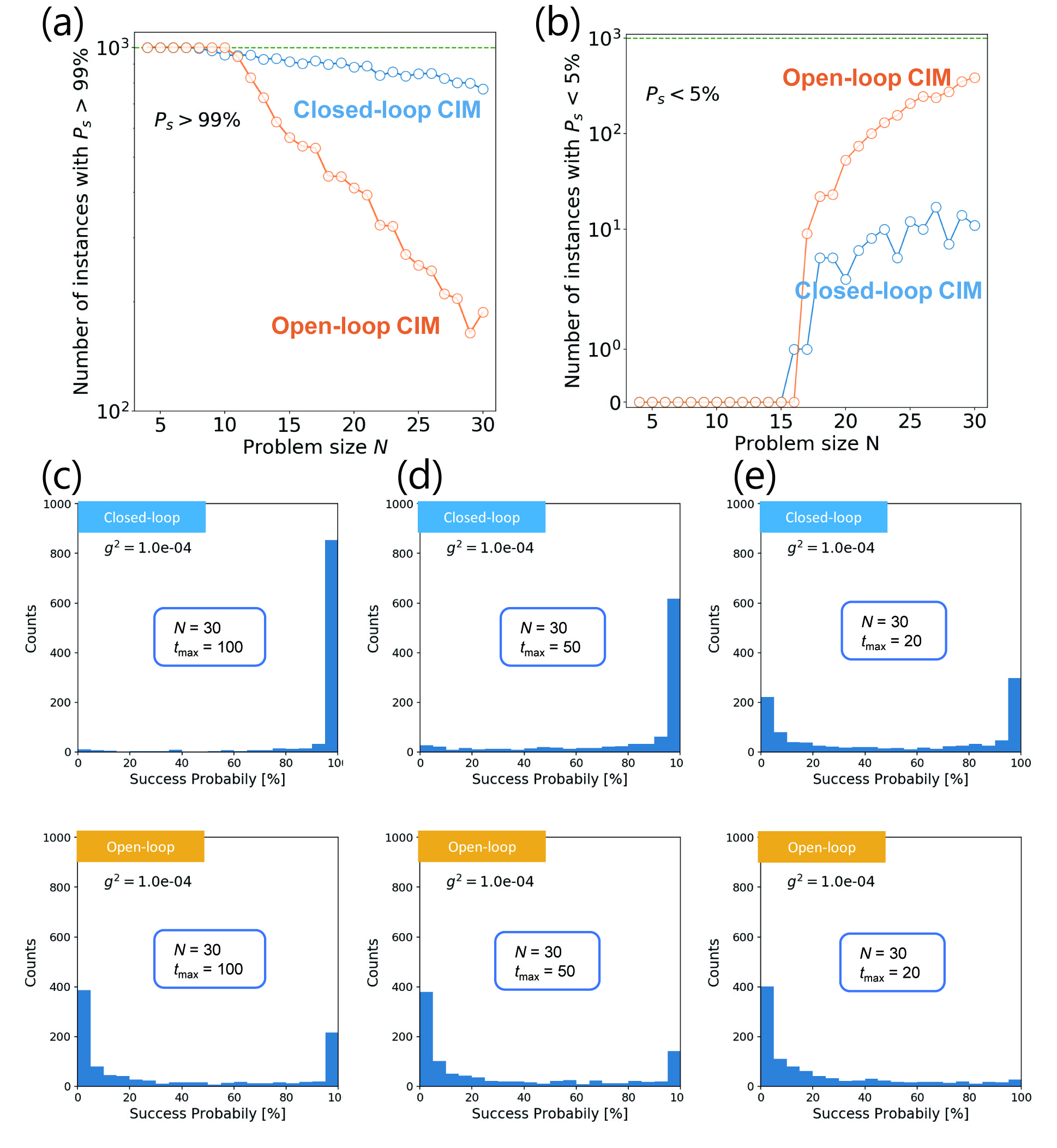}
	\caption{Performance comparison of the closed-loop CIM with error correction feedback and the open-loop CIM. The feedback parameters of the closed-loop CIM are $\beta = 0.05$, $\rho_a = \rho_p = 1.0$, $\alpha = 1.0$, $\pi = 0.2$, $\Delta = 1/5$, and $e(0)=1.0$. The parameters for the open-loop CIMs are $\beta = \rho_a = \rho_p = 0$, $p(t)=0.5(1+t/100)$, and $e(t)=1.0$. The ratio of a round trip time to a signal amplitude lifetime is ${\Delta t_c}/{\tau_{s}}=0.025$. (a) Number of instances with the success probability larger than 99\% vs. problem size. (b) Number of instances with the success probability lower than 5\% vs. problem size. (c)(d)(e) Histograms of success probabilities for the closed-loop and open-loop CIMs for the problems size of $N=30$. The saturation parameter is set to $g^2 = 10^{-4}$. The maximum computation time is $t_{max}=100$ for (c), $t_{max}=50$ for (d), and $t_{max}=20$ for (e), respectively.}
	\label{fig:size dependence}
\end{figure}

\subsection{Performance comparison against open-loop CIM}
To understand how the performance of a closed-loop CIM is compared to that of an open-loop CIM, we solve MAX-CUT problems with 21-level randomly chosen $J_{ik}$ and varying problem size $N=4 \sim 30$. A total of 1000 instances were generated for each problem size. Each problem instance is solved 100 times to evaluate the success probability. The maximum normalized computation time is set to $t_{max} = 100$. If the machine finds an exact solution at a certain time within $t_{max}$, we count it as a successful trial and evaluate the success probability by the total counts over 100 trials. Figure \ref{fig:size dependence} shows the performance of the closed-loop CIM with error correction feedback together with that of the open-loop CIM. The number of instances with a success probability higher than 99\% and lower than 5\% are plotted as a function of the problem size in figure \ref{fig:size dependence}(a) and (b), respectively. The feedback parameters of the closed-loop CIM are $\beta = 0.05$, $\rho = 1.0$, $\alpha = 1.0$, $\pi = 0.2$, $\Delta = 1/5$, and $e(0)=1.0$. The ratio of a round trip time to a signal field lifetime is $\Delta t_c / \tau_{s} = 0.025$. We study the open-loop CIM with the same Gaussian quantum model. We set the feedback mean-field strength $e_i(t) = e_i(0)=1.0$ constant for the open-loop CIM. The pump rate $p$ is linearly increased from $p=0.5$ at $t=0$ (below threshold) to $p=1.0$ at $t=100$ (above threshold). If the pump rate is abruptly switched on at $t=0$ from $p=0$ to $p=1.0$, the success probability is much worse.{\scriptsize $^{\cite{Hamerly2019}}$} 
\vspace{\baselineskip}

As shown in figure \ref{fig:size dependence}(a), the performance of the closed-loop CIM is superior to the open-loop CIM. The probability of finding the instances with a success probability higher than 99\% shown in figure \ref{fig:size dependence}(a), decreases much slower in the closed-loop CIM than the open-loop CIM. The probability of finding the instances with a success probability lower than 5\% increases dramatically in the case of the open-loop CIM but not in the closed-loop CIM, as shown in figure \ref{fig:size dependence}(b). The number of instances in the closed-loop CIM that give the success probability less than 5\% is an order of magnitude less than that in the open-loop CIM. The histogram of the success probabilities of the closed-loop CIM and the open-loop CIM are compared in figure \ref{fig:size dependence}(c), where the problem size is $N=30$, maximum computation time $t_{max}=100$ and the saturation parameter is $g^2=10^{-4}$. For the open-loop CIM, the instances are clearly separated into hard (success probability $\approx 0 \%$) or easy (success probability $\approx 100 \%$) instances. However, the closed-loop CIM can solve most of the instances with a high success probability. There are few instances ($<$ 1\%) with the success probability close to zero. Finally in figure \ref{fig:size dependence}(d)(e), where a maximum computation is decreased to $t_{max} = 50$ and $20$, it is shown that the superiority of the closed-loop CIM is preserved for shorter computation time. In figure \ref{fig:size dependence}(d) nearly 40\% of instances show the success probability close to zero for the open-loop CIM but the most of those instances can be solved with better success probability by the closed-loop CIM. Furthermore, in figure \ref{fig:size dependence}(e), the closed-loop CIM solves 30\% of the problem instances with nearly 100\% success probability but the open-loop CIM can solve only a few percent of the instances.

\begin{figure}[h]
	\centering
	\includegraphics[scale=0.18]{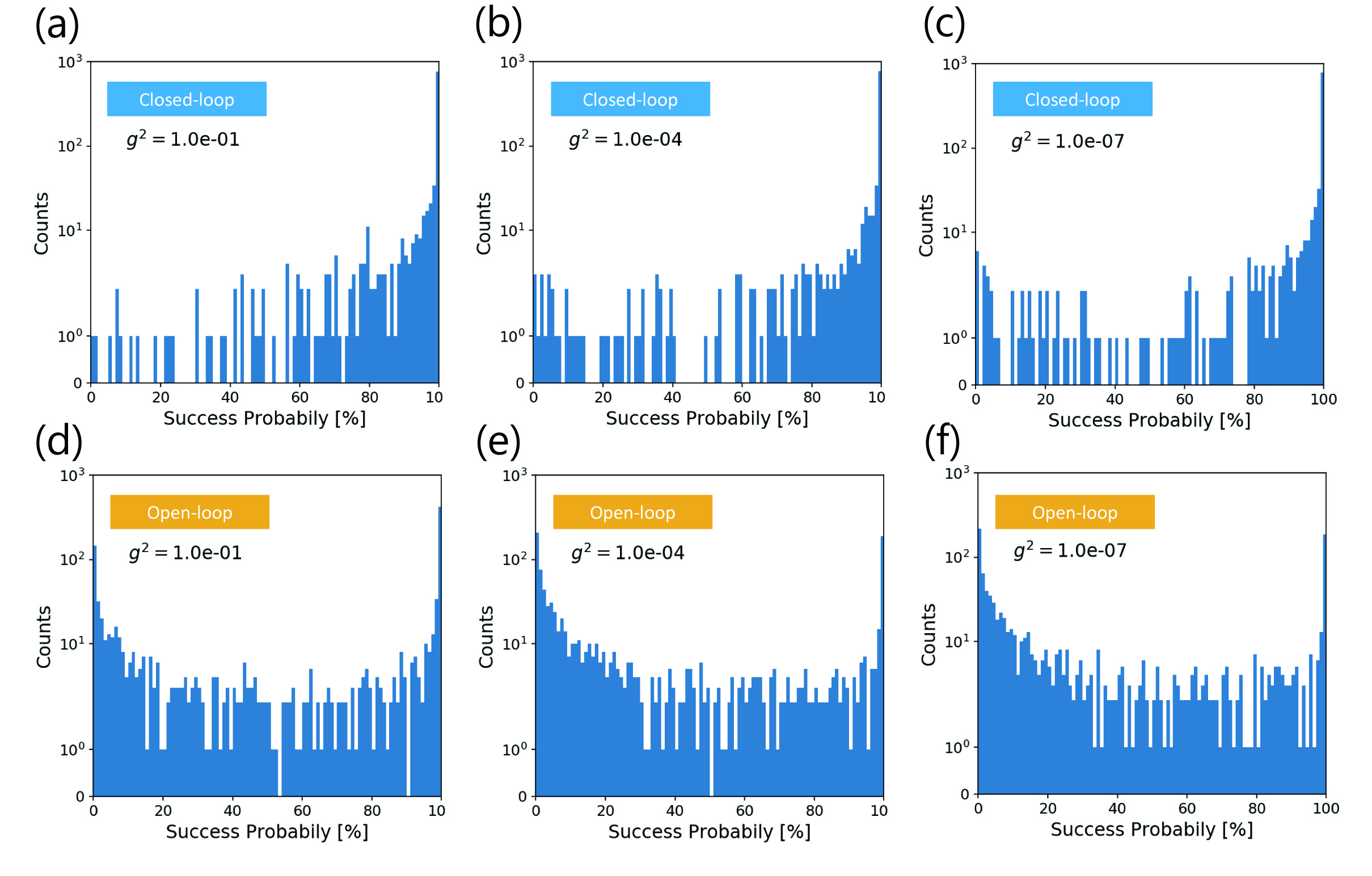}
	\caption{Success probability histogram of a closed-loop CIM, (a) $g^2 = 10^{-1}$, (b) $g^2 = 10^{-4}$, (c) $g^2 = 10^{-7}$, and open-loop CIM, (d) $g^2 = 10^{-1}$, (e) $g^2 = 10^{-4}$, (f) $g^2 = 10^{-7}$. A problem size $N=30$, round trip time $\Delta t_c = 0,025$ and number of round trips $N = 4 \times 10^3$ (or $t=100$).} 
	\label{fig:success probability}
\end{figure}

\subsection{Saturation parameter dependence}
The saturation parameter $g^2$ determines an inverse photon number at above threshold by ${\langle \hat{n} \rangle} \simeq \frac{(p - 1)}{g^2}$. If $g^2$ is much smaller than one, the coherent amplitude in DOPO is much larger than the vacuum fluctuation ${\langle \Delta \hat{X}^2 \rangle}^{1/2}_{SQL} = 1/\sqrt{2}$, so that an oscillator in the limit of $g^2 \rightarrow 0$ is considered as a classical oscillator. On the other hand, if $g^2$ is increased close to one, the coherent amplitude at above threshold is comparative to the vacuum  fluctuation, so that such an oscillator is considered as an oscillator in deep quantum regime (see Appendix C). Indeed, it is shown in ref. {\cite{Yamamura2017}} that a Schr\"{o}dinger cat-like state exists in an DOPO with $g^{2} \simeq 1$ at above threshold.
\vspace{\baselineskip}

In figure \ref{fig:success probability}(a)-(c), we show the success probability histogram for 1000 randomly generated problem instances against $g^2 = 10^{-1}$, $10^{-4}$ and $10^{-7}$. As the OPO in a closed-loop CIM increases its quanntumness, the success probability is greatly improved particularly for hard instances. When $g^2 = 10^{-7}$, about ten problem instances cannot be solved in $4 \times 10^3$ round trips (or 100 signal field lifetime). However, most of those hard instances with a success probability lower than 10\% are solved  when $g^2 = 10^{-1}$. In Figure \ref{fig:success probability}(d)-(f), we show the success probability for an open-loop CIM against $g^2 = 10^{-1}$, $10^{-4}$ and $10^{-7}$. The success probability is improved with increasing $g^2$ value, but the improvement is smaller than the closed-loop CIM.

\subsection{Random sampling in the closed-loop CIM}
In the previous section, we have shown the improved performance of the closed-loop CIM over the open-loop CIM, where the probability of finding one of the ground states in a single trial is evaluated. Here, we investigate how the proposed closed-loop CIM samples ground states as well as low-energy excited states for a given problem instance. In order to see the efficient and fair sampling performance of the machine, the probability of finding not only a specific energy state but also that for all degenerate states with a same energy are studied. We chose a particular problem instance that has the largest number of degenerate ground states from randomly generated 1000 instances. 
\vspace{\baselineskip}

\begin{figure}[h]
	\centering
	\includegraphics[scale=0.12]{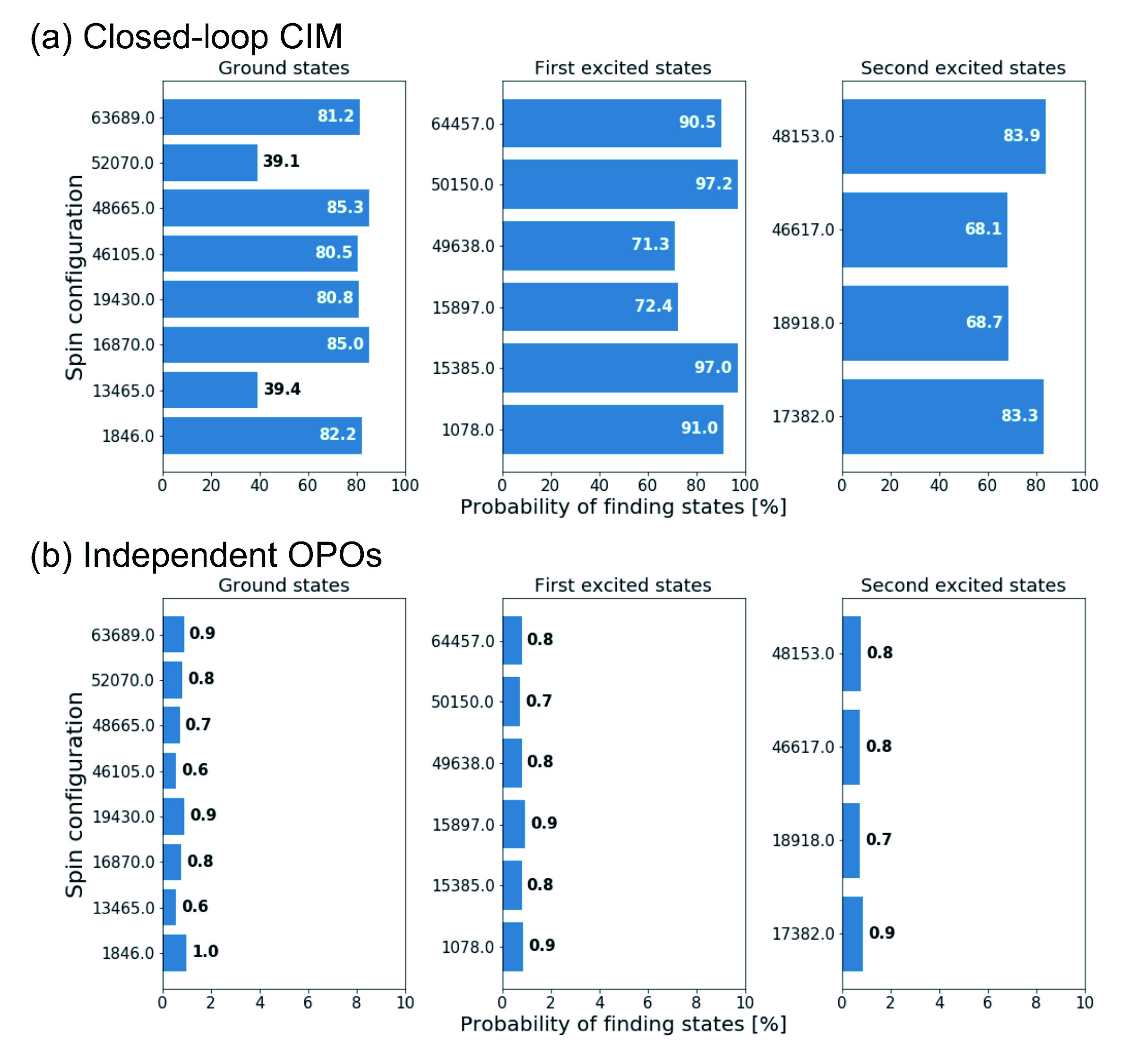}
	\caption{Sampling property of (a) closed-loop CIM and (b)uncoupled DOPOs as a function of the excess energy $\mathcal{E}$. (i) The probability of finding the specific energy states by the two systems. (ii) The theoretical Boltzmann distribution at an effective temperature of $T_{eff}=1.51$, and $T_{eff}\simeq 1000$, and (iii) the density of state for a given problem instance. All probabilities are normalized to unity. Inset shows the $D_{KL}$ as a function of the $T_{eff}$.}
	\label{fig:sampling-energy}
\end{figure}

The selected instance has a problem size of $N=16$ and has eight degenerate ground states, six degenerate first-excited states, and four degenerate second-excited states. The energy difference between each energy state is 0.2, which corresponds to an energy difference of a single spin flip and a minimum weight of 0.1. We solve this problem instance $10^4$ times to evaluate the sampling performance. The saturation parameter is $g^2=10^{-4}$ and the feedback parameters of the closed-loop CIM are $\alpha = \rho_{a} = 1.0 \times g^2$, $\beta = 0.05/\alpha$, $\pi = 0.2$, $\rho_{p} = 1.0$, $\Delta = 1/5$, and $e(0)=1.0$. The ratio of a round trip time to a signal field lifetime is set $\Delta t_{c} / \tau_{s} = 0.025$. The time interval $\Delta t$ in the numerical integration of equations \ref{eq:mu-equation},\ref{eq:val-equation},\ref{eq:error-equation} is identical to the round trip time of $\Delta t_c = 0.025$. The maximum computation time is $t_{max} = 100$, which indicates that there are $4 \times 10^{3}$ sampling events in a single trial. For comparison, we simulate independent DOPOs with the same Gaussian quantum model, but set the feedback parameters of $e(t)=0$ to cut off mutual coupling among DOPOs. Note that there still exists the measurement-induced shift of the mean-field $\mu_i$ and reduction of the variance $\sigma_i$. The pump rate $p=\pi(t)$ is linearly increased from $\pi(0)=1.5$ (below the threshold) to $\pi(100)=2.5$ (above the threshold). Note that the solitary DOPO threshold pump rate is $1+j = 2$.
\vspace{\baselineskip}

Figure \ref{fig:sampling-energy} shows the probabilities $P_{CIM}(\mathcal{E})$ of sampling a specific energy state by the closed-loop CIM (figure \ref{fig:sampling-energy}(a)) and $P_{DOPO} (\mathcal{E})$ by the independent DOPOs (figure \ref{fig:sampling-energy}(b)) vs. excess energy of $\mathcal{E}=\mathcal{E}_{Ising}-\mathcal{E}_{ground}$ measured from the ground state energy. The probability $P_{CIM}(\mathcal{E})$ is evaluated by averaging out the individual probability distribution of the $10^4$ trials. Each distribution for a trial is obtained by normalizing the histogram that represents how many times the machine samples a specific energy state at $\mathcal{E}_i$. The probability $P_{CIM}(\mathcal{E})$ is favorably compared to the theoretical Boltzmann distribution as shown in figure \ref{fig:sampling-energy}(a). The Boltzmann distribution at an effective temperature of $T_{eff}$ is given by
\begin{eqnarray}
	P_{Boltzmann}(\mathcal{E}_i)&=& D(\mathcal{E}_i) \times \frac{1}{\mathcal{Z}}\exp \left(-\frac{\mathcal{E}_i}{T_{eff}} \right) \\
	\mathcal{Z}&=&\sum_{i}{D(\mathcal{E}_i) \exp \left(-\frac{\mathcal{E}_i}{T_{eff}} \right)}\Delta \mathcal{E}.
\end{eqnarray}
Here $D(\mathcal{E}_i)$ is the density of state for the given problem instance and obtained by a brute force search. $\mathcal{E}_i$ represents the excess energy of the i-th bin of the histograms, $\Delta \mathcal{E}$ is the energy width of the histogram bin. The bin width of $\Delta \mathcal{E}$ is 0.2 in figure\ref{fig:sampling-energy}. The effective temperature $T_{eff}$ is estimated by minimizing Kullback-Leibler (KL) divergence $D_{KL}$ between the CIM probability distribution $P_{CIM}(\mathcal{E})$ and the Boltzmann distribution $P_{Boltzmann}$ (See the inset of figure\ref{fig:sampling-energy}(a)). The KL divergence $D_{KL}$ between two probability distributions $\{ P_n \}$ and $\{ Q_n \}$ is defined by
\begin{equation}
	D_{KL}(P\|Q)=\sum_{n}{P_n}\log \left( \frac{P_n}{Q_n} \right)
\end{equation}
Here we choose the Boltzmann distribution as $\{ P_n \}$ and the CIM distribution as $\{ Q_n \}$. The observed probability distributions $P_{CIM}(\mathcal{E}_i)$ of the closed-loop CIM is well matched with the Boltzmann distribution $P_{Bolzmann}(\mathcal{E}_i)$ at the fitted effective temperature of $T_{eff}=1.51$. On the other hand, the system of independent DOPOs possesses a much higher temperature ($T_{eff} \approx 1000$), so that the sampling is a truly random process.
\vspace{\baselineskip}

\begin{figure}[h]
	\centering
	\includegraphics[scale=0.15]{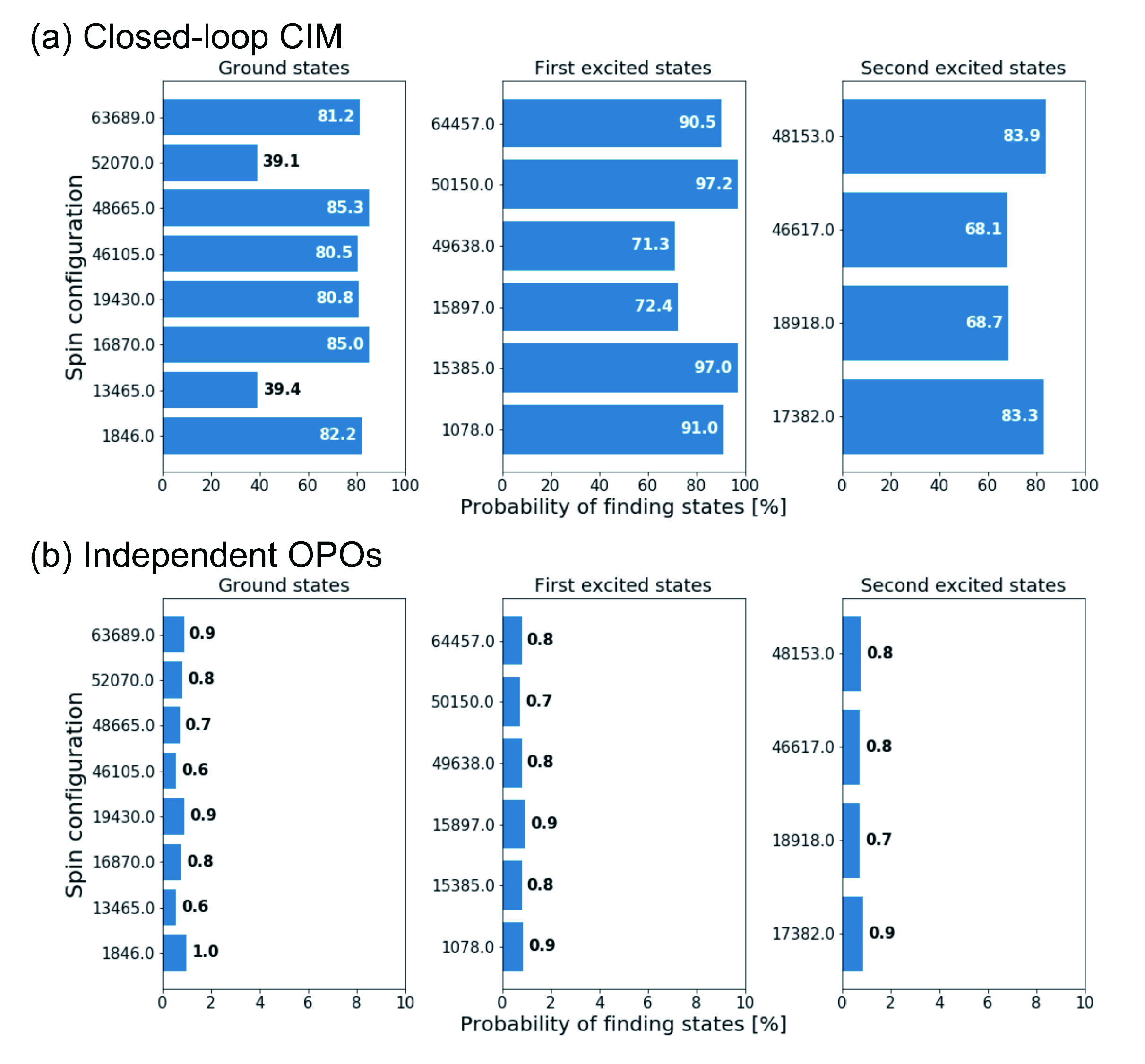}
	\caption{Sampling performance comparison for the closed-loop CIM and uncoupled DOPOs. (a) Probability of finding each state in the eight degenerate ground states, six degenerate first excited states and four degenerate second excited states by the closed-loop CIMs. (b) Those by independent DOPOs without Ising coupling. The labels for the vertical axis represent the number of a specific spin configuration of each degenerate state in decimal representation. The horizontal axis shows the probability of finding each state in a single trial.}
	\label{fig:sampling-state}
\end{figure}

Figure \ref{fig:sampling-state} shows how the closed-loop CIM samples the degenerate states in the lowest three energy states in a single run. The sampling performance of the closed-loop CIM is shown in figure \ref{fig:sampling-state}(a) while that of $N=16$ independent DOPOs is shown in figure \ref{fig:sampling-state}(b). All degenerate states are found for the closed-loop CIM with a higher probability than 60\% except for two complimentary ground states with the spin configurations of $13465$ and $52070$. Even for these hard states to sample, the success probabilities by a single run are higher than 35\%, indicating that three independent trials would be enough to pick up all {\scriptsize $^{\cite{Hamerly2019}}$} of the ground, first excited and second excited states. On the other hand, the corresponding probability for independent DOPOs is much lower, that is, of about $0.8 \sim 0.9\%$. This value is about seven times lower than a simple estimate $\approx$ 6\% of random guessing ($4\times10^3$ random sampling against $2^{16}$ states). This decrease in the probability for independent DOPOs is caused by the fact that the response time of each DOPO is longer than the sampling period, leading to a lower effective sampling rate. If the pump rate $p=\pi(t)$ is linearly increased from $\pi(0)=1.9$ (a little below the threshold) to $\pi(100)=2.5$ (a little above the threshold), the DOPOs evolve more slowly and result in an even lower probability of $0.3 \sim 0.4\%$.
\vspace{\baselineskip}

The simulation result of the closed-loop CIM shown in figures \ref{fig:sampling-energy} and \ref{fig:sampling-state} are obtained for a target intensity of $\alpha = \rho_a = 1.0 \times g^2$. In this parameter condition, the feedback mean-field $e_i(t)$ is modulated to stabilize the DOPO mean-field amplitude $\mu_{i}$ around $1$, where the amplitude of the quantum fluctuation is comparable to the mean-field. If we set the feedback parameters of $\alpha = \rho_{a} = 1.0$, the DOPO mean-field amplitude is stabilized around $\mu_{i}=1/g$, which is two orders magnitude larger than the amplitude of the vacuum fluctuation. The effective temperature $T_{eff}$ for the feedback parameters of $\alpha = \rho_{a} = 1.0$ is decreased to $T_{eff}=0.34$ and the sampling efficiency (probability) of finding the two ground states of $13465$ and $52070$ and the second-excited states are decreased. These results indicate that a closed-loop CIM can realize efficient random sampling of degenerate ground states and low-energy excited states by adjusting the effective temperature $T_{eff}$ of the machine through the feedback parameter of $\alpha$.

\subsection{Scaling to larger problem size}

To see how the performance of a closed-loop CIM scales to larger problem size, we solved the Sherrington-Kirkpatrick (SK) spin-glass model on a fully connected complete graph with $J_{ik} = \pm1$.{\scriptsize $^{\cite{Hamerly2019,Leleu2019}}$}  For each problem size $N=100 \sim 300$, 10 randomly generated SK spin-glass instances are solved by the closed-loop CIM. Each problem instance is solved 20 times to evaluate the success probability $\langle P_s \rangle$ of finding a ground state. The saturation parameter is set to $g^2 = 10^{-4}$.

\begin{center}
	\begin{table}[h]
		\caption{Averaged success probability $\langle P_s \rangle$ of finding a ground state after $t_{max}$ computation time and the median of the time-to-solution $t_s$ for the closed-loop CIM. The saturation parameter is set to $g^2 = 10^{-4}$. The time interval ${\Delta t}_c$ and the feedback parameters for $t_{max}$=$10^2$, $10^3$, and $10^4$ are the same as in figure \ref{fig:size dependence}. In the lowest two rows, the feedback parameters are optimized to $\rho_a = 1.0$, $\rho_p = 0$, $\alpha = 1.0$, $\pi = 0.2$, $\Delta = 1/4$, and $e(0)=1.0$. The time intervals of $t_{max}=N^2/10$ and $t_{max}=N^2$ are ${\Delta t}_c = 2^{-5}$ and ${\Delta t}_c = 2^{-6}$, respectively. $\beta$ is time-dependent as described in ref. \cite{Leleu2019} with the parameters $\gamma = 5.5 \times 10^{-4}$ and $\tau = 600$.
		}
		\label{tab:Ps-TTS}
		\small 
		\scalebox{1.0}[1.0]{
			\begin{tabular}{c|c|c|c|c|c|c|c|c|c|c}
				\hline
				\multirow{2}{*}{\backslashbox{$t_{max}$}{N}} &
				\multicolumn{2}{c|}{100} &
				\multicolumn{2}{c|}{150} &
				\multicolumn{2}{c|}{200} &
				\multicolumn{2}{c|}{250} &
				\multicolumn{2}{c}{300} \\ \cline{2-11} 
				&
				$\langle P_s \rangle$ {[}\%{]} &
				$t_s$ &
				$\langle P_s \rangle${[}\%{]} &
				$t_s$ &
				$\langle P_s \rangle${[}\%{]} &
				$t_s$ &
				$\langle P_s \rangle${[}\%{]} &
				$t_s$ &
				$\langle P_s \rangle${[}\%{]} &
				$t_s$ \\ \hline
				\begin{tabular}[c]{@{}c@{}}$10^2$\\ \end{tabular} &
				56 &
				$6.3 \times 10^2$ &
				14 &
				$3.5 \times 10^3$ &
				3 &
				- &
				0.5 &
				- &
				0.5 &
				- \\
				\begin{tabular}[c]{@{}c@{}}$10^3$\\ \end{tabular} &
				76 &
				$1.8 \times 10^3$ &
				46 &
				$7.8 \times 10^3$ &
				18 &
				$6.0 \times 10^4$ &
				7 &
				- &
				2 &
				- \\
				\begin{tabular}[c]{@{}c@{}}$10^4$\\ \end{tabular} &
				89 &
				$1.6 \times 10^4$ &
				95 &
				$1.6 \times 10^4$ &
				45 &
				$9.3 \times 10^4$ &
				29 &
				$1.6 \times 10^5$ &
				7 &
				$9.0 \times 10^5$ \\ \hline
				\begin{tabular}[c]{@{}c@{}} $N^2/10$ \end{tabular} &
				93 &
				$1.5 \times 10^3$ &
				46 &
				$2.1 \times 10^4$ &
				37 &
				$4.7 \times 10^4$ &
				36 &
				$7.4 \times 10^4$ &
				22 &
				$1.9 \times 10^5$ \\
				\begin{tabular}[c]{@{}c@{}} $N^2$ \end{tabular} &
				94 &
				$1.5 \times 10^4$ &
				93 &
				$4.0 \times 10^4$ &
				88 &
				$7.1 \times 10^5$ &
				84 &
				$1.4 \times 10^5$ &
				90 &
				$1.8 \times 10^5$ \\ \hline
			\end{tabular}
		}
	\end{table}
\end{center}

In table \ref{tab:Ps-TTS}, we show the success probability $P_s$ of finding a ground state after computation time $t_{max}$ and the median of the time-to-solution  $t_{s}$ for $t_{max}$ = $10^2$, $10^3$, $10^4$, $N^2/10$, and $N^2$. The averaged success probability  $\langle P_s \rangle$ is obtained by averaging $P_s$ over 10 SK instances of each problem size $N$. The time-to-solution is the expected computation time required to find a ground state with 99\% probability with a single run. The time-to-solution $t_s$ of a given problem instance is estimated by $t_s = t_{max} \log (1-0.99)/\log(1-P_s)$. The success probability $\langle P_s \rangle$ decreases with increasing the problem size $N$ for fixed computation time $t_{max}$. The lowest two rows in table \ref{tab:Ps-TTS} show the simulation results obtained by adjusting $t_{max}$ as a function of the problem size and by using optimized feedback parameters.{\scriptsize $^{\cite{Leleu2019}}$} As shown in the lowest row in table \ref{tab:Ps-TTS}, the closed-loop CIM with a computation time $t_{max} = N^2$can solve the SK spin-glass instances  with a success probability nearly 90\% up to $N=300$. 
\vspace{\baselineskip}

\begin{figure}[h]
	\centering
	\includegraphics[scale=0.12]{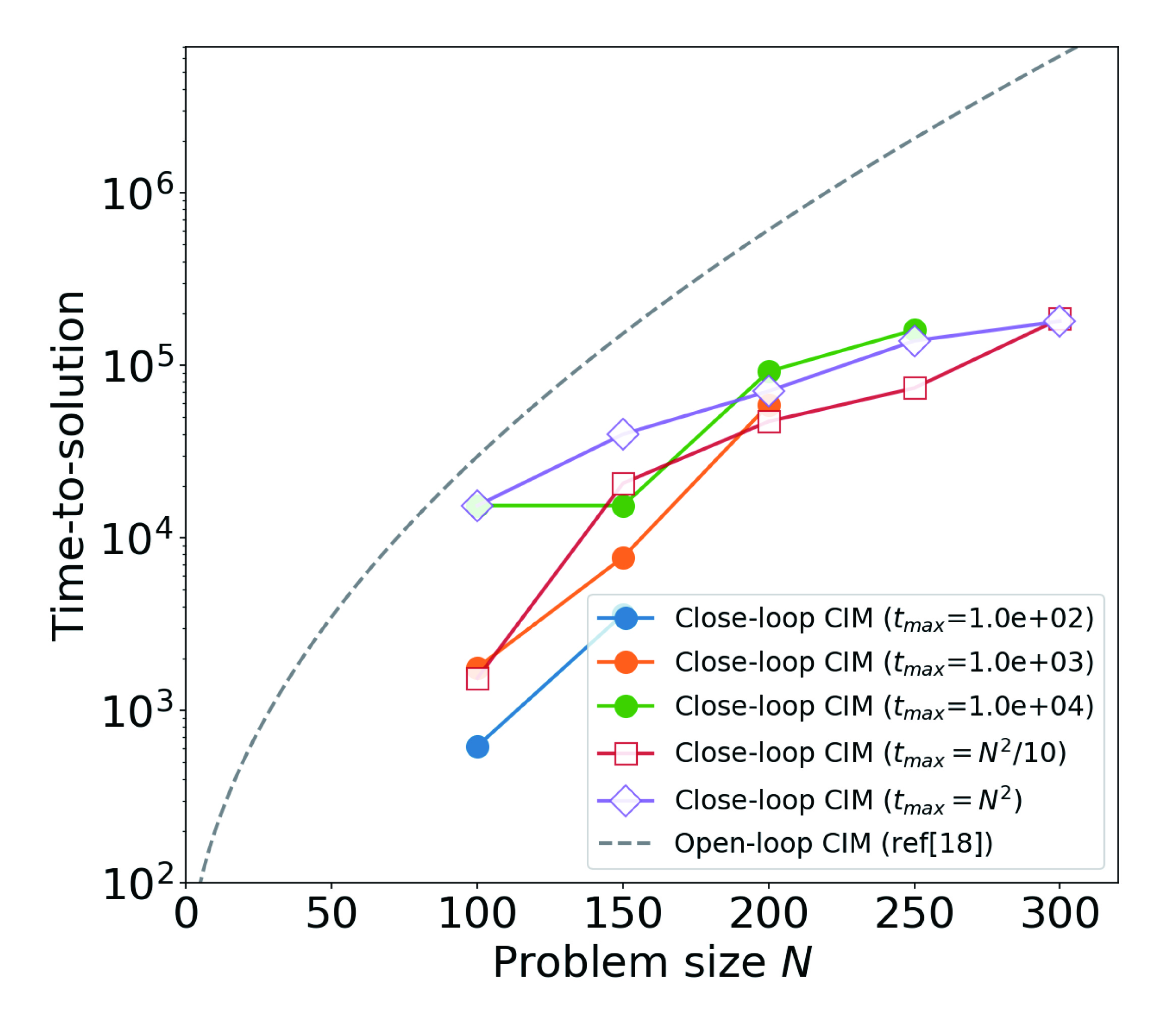}
	\caption{Scaling properties of the median of the time-to-solution $t_s$ for problem size of $N=100 \sim 300$. The data can be found in table \ref{tab:Ps-TTS} for the closed-loop CIM. The time-to-solution of closed-loop CIM  for the $t_{max}$ = $10^2$, $10^3$, $10^4$, $N^2/10$ and $N^2$ is normalized by signal amplitude lifetime $\tau_{s}$. The dashed line shows the optimal time-to-solution of the state-of-art open-loop CIM in the time unit normalized by round-trip time (see ref. \cite{Hamerly2019}).}
	\label{fig:large-Size}
\end{figure}

Figure \ref{fig:large-Size} plots the medians of the time-to-solution of the closed-loop CIM for $t_{max}$ = $10^2$, $10^3$, $10^4$, $N^2/10$ and $N^2$, in conjunction with the optimal time-to-solution of the state-of-art open-loop CIM{\scriptsize $^{\cite{Hamerly2019}}$}. It is difficult to compare directly with the properties of the open-loop CIM because the theoretical models are slightly different in reference,{\scriptsize $^{\cite{Hamerly2019}}$} but the closed-loop CIM shows similar or slightly better scaling properties to the state-of-art open-loop CIM. 

\section{Conclusion}
We have numerically studied the performance of the closed-loop CIM with error detection and correction feedback, in which amplitude squeezed states of DOPO pulses are repeatedly and weakly monitored by optical homodyne measurement and displaced by error correction feedback signals. A Gaussian quantum model, which is derived from the measurement-feedback CIM master equation using the Wigner representation for the field density operator, is used to simulate the dynamical behavior of mean-field amplitudes and variances. This approximate model is valid as far as a saturation parameter is small ($g^2\ll1$) and a signal field lifetime is much longer than a cavity round trip time (${\tau_{s}}/{\Delta t_c}\gg1$).
\vspace{\baselineskip}

The closed-loop CIM is expected to partially overcome the two drawbacks of the previously studied open-loop CIM: (1) exponentially increasing local minima trap a machine state as a problem size increases, and (2) mapping of a target Hamiltonian to a loss landscape fails due to DOPO amplitude heterogeneity. The above expectation is confirmed by comparing the performance of a closed-loop CIM to that of an open-loop CIM. Moreover, it is shown that the hopping behavior of a closed-loop CIM realizes efficient sampling of degenerate ground states and low-energy excited states, which is useful for various applications including the lead optimization for drug discovery.{\scriptsize $^{\cite{Sakaguchi2016}}$} The error detection and correction feedback mechanism allows a greater reduction of the unstable manifold dimension of states associated with low-energy excited states. Thus, the closed-loop CIM has the potential to sample rapidly from many local minima of low-energy states rapidly during a single trial.
\vspace{\baselineskip}

In a future publication, we will report on the performance of a closed-loop CIM in an opposite operating regime of a low-Q  cavity ($\Delta t_c / t_{s} \gtrsim 0.1$), which requires a new theoretical tool beyond the present Gaussian quantum model based on the Wigner stochastic differential equation.

\subsection*{Appendix A: Gaussian quantum model for measurement feedback coupling CIM}	
The master equation for the density operator of a solitary DOPO is 
\setcounter{equation}{0} \renewcommand{\theequation}{A\arabic{equation}}
\begin{equation}{\label{eq:density operator}}
\frac{d}{dt} \hat{\rho}=\frac{1}{i \hbar}\left[ \hat{\mathcal H}, \hat{\rho}\right]  + \sum_{j=1,2} \left( \left[ \hat{L}_j, \hat{\rho}\hat{L}^+_j\right] +h,c\right) , 
\end{equation}
where $\hat{\mathcal H} = i \hbar \frac{p}{2}(\hat{a}^{+2} - \hat{a}^2)$ is the parametric interaction Hamiltonian, and $\hat{L}_1 = \hat{a}$ and $\hat{L}_2 = \sqrt{\frac{g}{2}} \hat{a}^2$ are the projectors for linear loss and two photon loss, respectively. The stochastic differential equation (SDE) for the complex amplitude $\alpha$ of a solitary DOPO is obtained from the master equation (\ref{eq:density operator}) with Wigner expansion $W(\alpha)$ of $\hat{\rho}$ {\scriptsize $^{\cite{Wang2013}}$}{\scriptsize $^{\cite{Maruo2016}}$}
\begin{equation}{\label{eq:SDE}}
\frac{d}{dt} \alpha = - \alpha + p \alpha^* - g^2 |\alpha|^2 \alpha + \sqrt{\frac{1}{2} + g^2 |\alpha|^2} \xi_c, 
\end{equation}
where $\xi_c$ is a complex random variable satisfying $\left\langle \xi^*_c (t) \xi_c (t') \right\rangle = 2 \delta (t - t') $. As shown in figure \ref{fig:setup}, the coupling among DOPOs in MFB-CIM is introduced by optical homodyne measurement of in-phase amplitude of an external (measured) DOPO field and injection feedback to in-phase amplitude of an internal target DOPO field, while quadrature-phase amplitude is not measured and left uncoupled.
\vspace{\baselineskip}

From the Gaussian homodyne measurement theory{\scriptsize $^{\cite{Eisert2002}}$} applied to the measurement-feedback circuit shown in figure \ref{fig:setup}, the mean amplitude shift and the variance reduction induced by measurement action are expressed as
\begin{eqnarray}
{\label{eq:Gaussian 1}}
\langle \hat{X}' \rangle &=& \langle \hat{X} \rangle + \frac{\langle \Delta \hat{X} \Delta \hat{X}_R \rangle}{\frac{1}{2} + \langle :\Delta \hat{X}^2_R: \rangle} \delta, \\
{\label{eq:Gaussian 2}}
\langle :\Delta \hat{X}^{'2}: \rangle &=& \langle :\Delta \hat{X}^2: \rangle - \frac{{\langle \Delta \hat{X} \Delta \hat{X}_R \rangle}^2}{\frac{1}{2} + \langle :\Delta \hat{X}^2_R: \rangle}.
\end{eqnarray}
Here $\langle \hat{X} \rangle$ and $\langle \hat{X}' \rangle$ are the mean amplitudes of the internal field before and after the measurement, $\langle :\Delta \hat{X}^{'2}: \rangle$ and $\langle :\Delta \hat{X}^2: \rangle$ are the normally ordered variances of the internal field before and after the measurement, $\langle \Delta \hat{X} \Delta \hat{X}_R \rangle$ is the correlation function between internal (transmitted at XBS) and external (reflected at XBS) amplitudes, and $\delta = \tilde{X} - \langle \hat{X}_R \rangle$ is the difference of a measured value $\tilde{X}_R$ from the mean external amplitude. $\langle :\Delta\hat{X}^2:\rangle = \langle \Delta\hat{X}^2 \rangle - \frac{1}{2}$ is a normally ordered variance with vacuum fluctuation removed, so that the external amplitude variance is dependent only on the internal amplitude variance, that is, $\langle : \Delta\hat{X}^2_R:\rangle = R_B \langle :\Delta\hat{X}^2:\rangle$, where $R_B$ is the reflectivity of XBS. We assume a roundtrip time of a ring cavity $\Delta t_c$ is sufficiently shorter than a signal field lifetime, that is, $\Delta t_c \ll 1$. Therefore, background loss, measurement loss and parametric gain per roundtrip is small compared to one, which is an implicit assumption for deriving equation(\ref{eq:SDE}).
\vspace{\baselineskip}

The reflectivity $R_B$ of XBS is expressed conveniently as $R_B = j \Delta t_c$ and is sufficiently smaller than one where $j$ is a distributed outcoupling coefficient. Therefore, we can safely neglect $\langle :\Delta \hat{X}^2_R : \rangle$ in the denominator in equations(\ref{eq:Gaussian 1}) and (\ref{eq:Gaussian 2}) and the correlation function in the numerator of equations(\ref{eq:Gaussian 1}) and (\ref{eq:Gaussian 2}) is approximated by $\langle \Delta\hat{X} \Delta\hat{X}_R\rangle = \sqrt{R}_B \langle :\Delta\hat{X}^2:\rangle$. Under those approximations, equation(\ref{eq:Gaussian 2}) can be simplified as
\begin{equation}{\label{eq:Gaussian 3}}
\langle :\Delta \hat{X}^{'2}: \rangle = \langle :\Delta \hat{X}^2 : \rangle - 2j \Delta t_c {\langle : \Delta \hat{X}^2 : \rangle}^2.
\end{equation}
Note that the variance reduction is independent of an actual measurement result $\tilde{X}_R$ but is uniquely determined by the internal field variance $\langle :\Delta \hat{X}^2: \rangle$ and measurement strength $R_B = j \Delta t$. Under the same approximations, the external amplitude is expressed as $\hat{X}_R \simeq \sqrt{R_B} \langle \hat{X} \rangle - \sqrt{2(1 - R_B)} f_1 $, where $f_1$ is a real part of the complex random variable representing vacuum fluctuation incident upon XBS (see figure \ref{fig:setup}). The difference $\delta$ of a measured value from the mean amplitude, $\delta = -\sqrt{2 (1 - R_B)} f_1$, is related to the real number Gaussian random variable $w$ in equation(\ref{eq:mu-equation}) of the text by $w = \sqrt{\frac{2}{\Delta t_c}} \delta$, thus $\langle w(t) w(t') \rangle = \delta (t - t')$. Therefore, equation(\ref{eq:Gaussian 1}) can be simplified as
\begin{equation}{\label{eq:Gaussian 4}}
\langle \hat{X}' \rangle = \langle \hat{X} \rangle + \sqrt{2j} \langle : \Delta \hat{X}^2 : \rangle w \Delta t_c.
\end{equation}

If we substitute equations (\ref{eq:Gaussian 3}) and (\ref{eq:Gaussian 4}) into the real part of equation(\ref{eq:SDE}), we obtain the Winger SDE in Gaussian approximation for the mean amplitude and variance:{\scriptsize $^{\cite{Inui}}$}
\begin{eqnarray}
{\label{eq:mu-equation A}}
\frac{d}{dt} \mu_i &=& \left[ - \left( 1 + j \right) + p - g^2 \mu^2_i \right] \mu_i + \sum_{k}{J_{i k} j \left( \mu_k + \sqrt{\frac{1}{4j}} w_k \right)} + \sqrt{j} \langle :\Delta \hat{X^2_i}: \rangle w_i,\\
{\label{eq:val-equation A}}
\frac{d}{dt} \sigma_i &=& 2 \left[ - \left( 1 + j \right) + p - 3 g^2 \mu^2_i \right] \sigma_i - 2j {\left( \sigma_i -1/2 \right)}^2 + \left[ \left( 1 + j \right) + 2 g^2 \mu^2_i \right].
\end{eqnarray}
Here $\mu_i = \langle \hat{X}_i \rangle / 2 $ and $\sigma_i = \langle \Delta \hat{X}^2_i \rangle$. Equations (\ref{eq:mu-equation A}) and (\ref{eq:val-equation A}) are identical to equations (1) and (2) in the main text.

\subsection*{Appendix B: Amplitude heterogeneity in open-loop and closed-loop CIM}  
\setcounter{equation}{0} \renewcommand{\theequation}{B\arabic{equation}}

An open-loop CIM does not have any mechanism to remove amplitude heterogeneity among DOPOs. In such a case, the mapping protocol from an Ising Hamiltonian to a loss landscape breaks down.{\scriptsize $^{\cite{Wang2013, Kalinin2018, Leleu2017}}$} Figure \ref{fig:mean amplitude evolution}(a) shows the evolution of mean amplitude, $\mu(t)$ for a $N=16$ MAX=CUT problem with 21-level $J_{ik}$ weights. Steady state amplitudes at above threshold are highly inhomogeneous due to the lack of amplitude stabilization feedback. The variances $\langle \Delta {\hat{X}}^2 \rangle$ and $\langle \Delta {\hat{P}}^2 \rangle$ are also  inhomogeneous in spite of uniform pump rate for all DOPOs. One DOPO approaches a coherent state with $\langle \Delta {\hat{X}}^2 \rangle = \langle \Delta {\hat{P}}^2 \rangle = 1/2$, while the other DOPO remains at a squeezed vacuum state with $\langle \Delta {\hat{X}}^2 \rangle> 1/2$ and $ \langle \Delta {\hat{P}}^2 \rangle < 1/2$.
\vspace{\baselineskip}

\setcounter{figure}{0} \renewcommand{\thefigure}{B\arabic{figure}}
\begin{figure}[h]
	\centering
	\includegraphics[scale=0.15]{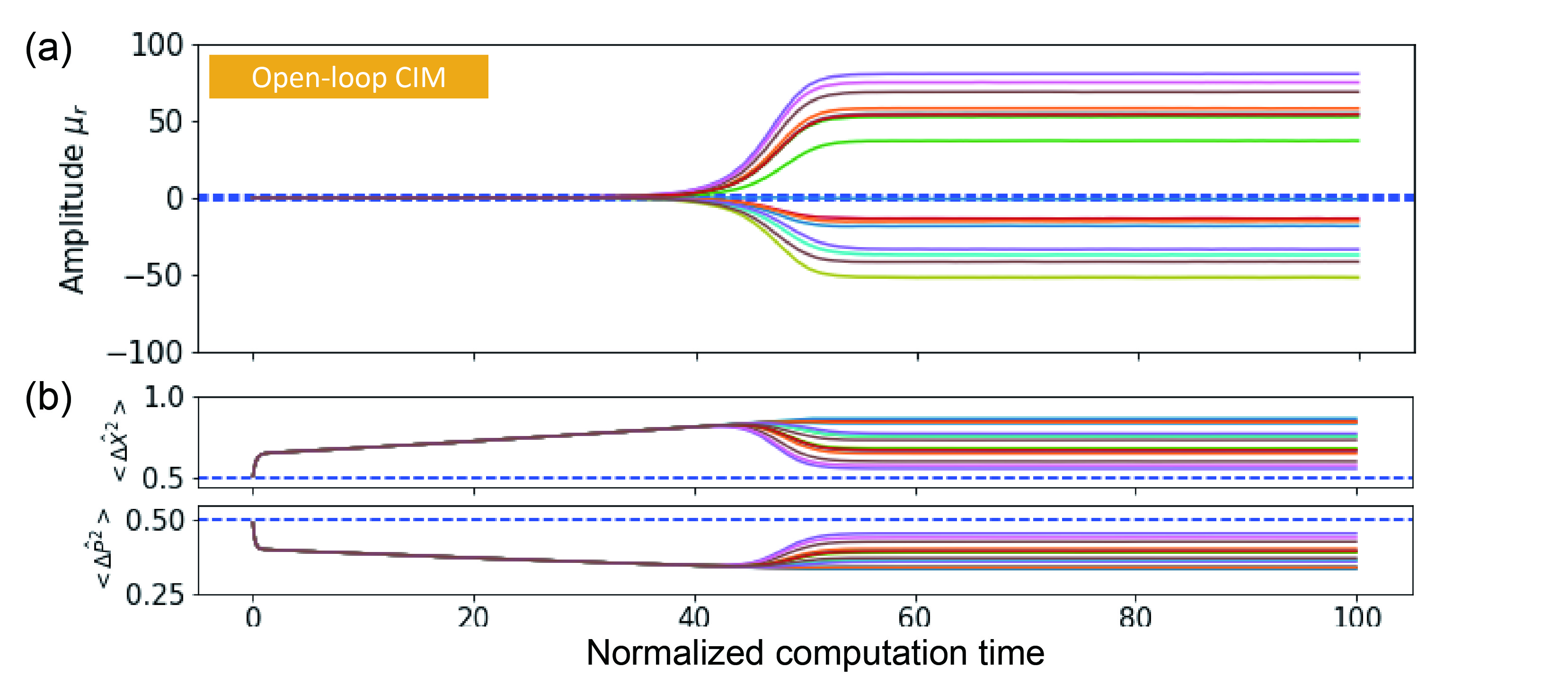}
	\caption{(a) Evolution of mean amplitude $\mu(t)$, (b) in-phase variance $\langle \Delta {\hat{X}}^2 \rangle$ and quadrature-phase variance $\langle \Delta {\hat{P}}^2 \rangle$ of an open-loop CIM. The same problem instance is used as figure \ref{fig:time response} in the main text.}.
	\label{fig:mean amplitude evolution}
\end{figure}

On the other hand, the evolution of mean amplitudes $\mu(t)$ and the variances $\langle \Delta {\hat{X}}^2 \rangle$ and $\langle \Delta {\hat{P}}^2 \rangle$ for a closed-loop CIM for the same problem instance in figure \ref{fig:time response}(c), (f) and (g) show that majority of DOPOs keep stabilized amplitudes and variances except for a few spin flipping DOPOs at specific timing. Because of the dynamical nature of a closed-loop CIM with varying target intensity $a(t)$, the amplitude of DOPOs never converge to a certain value but constantly evolve. Nevertheless, the amplitude heterogeneity is improved from an open-loop CIM.

\subsection*{Appendix C: Saturation parameter of optical parametric oscillators}
The Hamiltonian of a solitary degenerate optical oscillator is expressed as  
\setcounter{equation}{0} \renewcommand{\theequation}{C\arabic{equation}}
\begin{equation}{\label{eq:Hamiltonian}}
\hat{\mathcal H} = \hat{\mathcal H}_{free} + \hat{\mathcal H}_{int} + \hat{\mathcal H}_{pump} + \hat{\mathcal H}_{SR}, 
\end{equation}
where a free field Hamiltonian $\hat{\mathcal H}_{free}$, parametric interaction Hamiltonian $\hat{\mathcal H}_{int}$, pump Hamiltonian $\hat{\mathcal H}_{pump}$ and system-reservoir interaction Hamiltonian $\hat{\mathcal H}_{SR}$ are
\begin{eqnarray}
{\label{eq:free}}
\hat{\mathcal H}_{free} &=& \hbar \omega_s \hat{a}^+_s \hat{a}_s + \hbar \omega_p \hat{a}^+_p \hat{a}_p,\\
{\label{eq:int}}
\hat{\mathcal H}_{int} &=& i \frac{\hbar \kappa}{2} \left( \hat{a}^{+2}_s \hat{a}_p - \hat{a}^+_p \hat{a}^2_s \right),\\
{\label{eq:pump}}
\hat{\mathcal H}_{pump} &=& i \hbar \mathcal{E} \left( \hat{a}^+_p - \hat{a}_p \right),\\
{\label{eq:SR}}
\hat{\mathcal H}_{SR} &=& \hbar \left( \hat{a}_s \hat{\Gamma}^+_{R_s} + \hat{\Gamma}_{R_s} \hat{a}^+_s + \hat{a}_p \hat{\Gamma}^+_{R_p} + \hat{\Gamma}_{R_p} \hat{a}^+_p \right),
\end{eqnarray}
The master equation for a density operator for combined signal-pump-reservoir system, $\frac{d}{dt} \hat{\rho}_{total} = \frac{1}{i \hbar} [\hat{\mathcal H}, \hat{\rho}_{total}]$ can be simplified by eliminating the reservoir operators $\hat{\Gamma}_{R_s}$ and $\hat{\Gamma}_{R_p}$ by the Wigner-Weisskopt approximation:
\begin{equation}{\label{eq:Wigner-Weisskopt}}
\frac{d}{dt} \hat{\rho}_{s-p} =  \frac{1}{i \hbar} \left[ \hat{\mathcal H}_{free} + \hat{\mathcal H}_{int} + \hat{\mathcal H}_{pump}, \hat{\rho}_{s-p} \right] + \gamma_s \left( \hat{a}_s \hat{\rho}_{s-p} \hat{a}^+_s - \frac{1}{2} \left[ \hat{a}^+_s \hat{a}_s, \hat{\rho}_{s-p}\right] \right) +  \gamma_p \left( \hat{a}_p \hat{\rho}_{s-p} \hat{a}^+_p - \frac{1}{2} \left[ \hat{a}^+_p \hat{a}_p, \hat{\rho}_{s-p}\right] \right),
\end{equation}
where $\hat{\rho}_{s-p}$ is the combined system-pump density operator after tracing out the reservoir coordinates, and $\gamma_s$ and $\gamma_p$ are the Fermi's golden rule decay rates of signal and pump fields. If we further eliminate the pump field operator by assuming $\gamma_p \gg \gamma_s$, we can obtain the master equation (\ref{eq:density operator}) for the signal field operator and the Wigner stochastic differential equation (\ref{eq:SDE}).\\
\vspace{\baselineskip}

The saturation parameter $g^2$ is defined as
\begin{equation}{\label{eq:saturation parameter}}
g^2 = \frac{\kappa^2}{2 \gamma_s \gamma_p},
\end{equation}
which is much smaller than one for a OPO system with small parametric interaction ($\kappa$) and large dissipation rates ($\gamma_s$, $\gamma_p$) but can be made close to one by increasing $\kappa$ and decreasing $\gamma_s$ and $\gamma_p$. The threshold pump rate  $\mathcal{E}_{th}$ is expressed as
\begin{equation}{\label{eq:threshold pump rate}}
\mathcal{E}_{th} = \frac{\gamma_s \gamma_p}{\kappa} = \frac{\kappa}{2g^2}.
\end{equation}
The normalized pump rate $p$ in equation (\ref{eq:density operator}) is defined as $p = \mathcal{E}/\mathcal{E}_{th}$. By introducing a normalized in-phase amplitude by $c = Re (g \alpha)$ in equation (\ref{eq:SDE}), we obtain the following W-SDE:
\begin{equation}{\label{eq:W-SDE}}
\frac{d}{dt} c = \left[ -1 +p - c^2 \right]c +g \sqrt {\frac{1}{2} + c^2} \xi_r,
\end{equation}
where $\xi_r$ is a real part of $\xi_c$ in equation (\ref{eq:SDE}) and satisfies $\left[ \xi_r(x), \xi_r(x')\right] = \delta (x - t') $. As can be seen from equation (\ref{eq:W-SDE}), the quantum noise $\xi_r$ is neglected against the normalized amplitude $c$ and an OPO behaves as a classical (noise free) oscillator in the limit of $g^2 \rightarrow 0$.
\vspace{\baselineskip}

\setcounter{figure}{0} \renewcommand{\thefigure}{C\arabic{figure}}
\begin{figure}[h]
	\centering
	\includegraphics[scale=0.15]{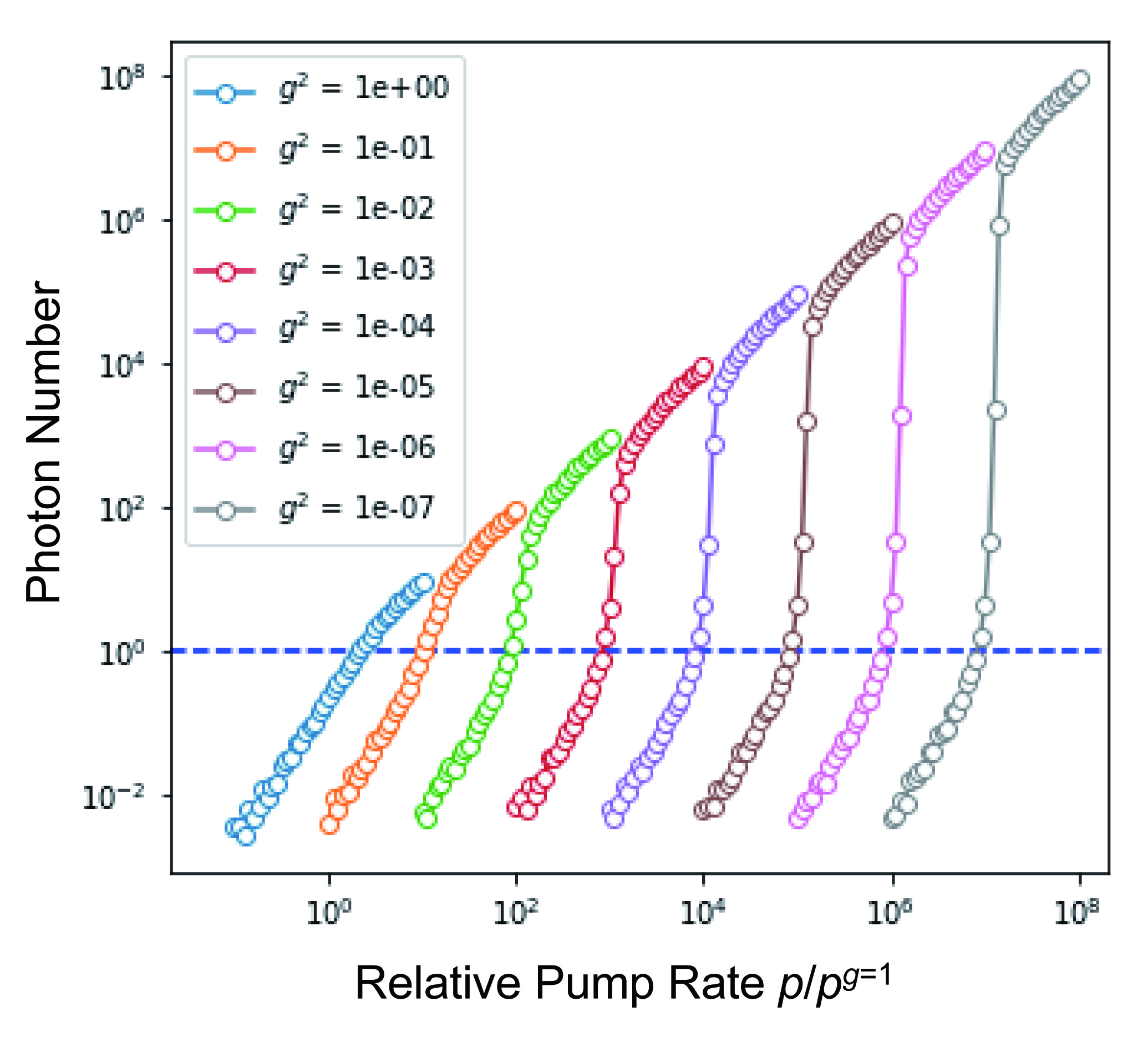}
	\caption{Average photon number $\left\langle \hat{n} \right\rangle$ vs. relative pump rate $p/p_{th}(g^2 = 1)$, where $p_{th}(g^{2} = 1)$ is a threshold pump rate  for $g^{2} = 1$.}
	\label{fig:photon number}
\end{figure}

Figure \ref{fig:photon number} shows the mean photon number $\left\langle \hat{n} \right\rangle = \mu^2$ vs. normalized pump rate $p/p_{th}(g^2 = 1)$ for various values of $g^2$. When $g^{2} = 1$, the input-output relation features a so-called ``thresholdless-like'' behavior as seen in a laser with a spontaneous emission coupling coefficient $\beta = 1$.{\scriptsize $^{\cite{Bjork1994}}$} As $g^2$ decreases, the threshold pump rate increases as shown in equation (\ref{eq:threshold pump rate}) and the average photon number at above threshold, also increases as $\left\langle \hat{n} \right\rangle = (p - 1)/g^2$.

\medskip
\textbf{Acknowledgements} \par 
The authors wish to thank the useful discussions with Hideo Mabuchi, Surya Ganguli, Zoltán Toroczkai, Peter Drummond, Margaret Reid.

\medskip
\textbf{Conflict of Interest} \par 
The authors have no conflict of interest, financial or otherwise.

\medskip

%

\bibliographystyle{plain}

\end{document}